\documentclass{article}

\usepackage{epubtk}
\usepackage{graphicx}
\usepackage{amssymb}
\usepackage{mathrsfs}

\begin{document}

\def\scri{$\mathscr{I}^+$}

\title{Critical phenomena in gravitational collapse}

\author{
\epubtkAuthorData{Carsten Gundlach}{
School of Mathematics, University of Southampton, \\
Southampton SO17 1BJ, UK}{
cjg@soton.ac.uk}{
http://www.soton.ac.uk/~cjg}
\and
\epubtkAuthorData{Jos\'e M. Mart\'{\i}n-Garc\'{\i}a}{
Instituto de Estructura de la Materia,
Centro de F\'{\i}sica Miguel A. Catal\'an, \\
C.S.I.C.,
Serrano 123, 28006 Madrid, Spain}{
jmm@iem.cfmac.csic.es}{ 
http://metric.iem.csic.es/Martin-Garcia/}
}

\date{}
\maketitle

\begin{abstract}

As first discovered by Choptuik, the black hole threshold in the space
of initial data for general relativity shows both surprising structure
and surprising simplicity. Universality, power-law scaling of the
black hole mass, and scale echoing have given rise to the term
``critical phenomena''. They are explained by the existence of exact
solutions which are attractors within the black hole threshold, that
is, attractors of codimension one in phase space, and which are
typically self-similar. Critical phenomena give a natural route from
smooth initial data to arbitrarily large curvatures visible from
infinity, and are therefore likely to be relevant for cosmic
censorship, quantum gravity, astrophysics, and our general
understanding of the dynamics of general relativity.

\end{abstract}

\tableofcontents


\section{Introduction}


\subsection{Overview of the subject}

Take generic initial data in general relativity, adjust
any one parameter $p$ of the initial data to the threshold of black
hole formation, and compare the resulting spacetimes as a function of
$p$. In many situations, the following {\em critical phenomena} are
then observed:
\begin{itemize}

\item Near the threshold, black holes with arbitrarily small masses can be
created, and the black hole mass scales as 
\begin{equation}
\label{eq1}
M\propto (p-p_*)^\gamma,
\end{equation}
where $p$ parameterises the initial data and black holes form for
$p>p_*$. 

\item The {\em critical exponent} $\gamma$ is universal with respect to
initial data, that is, independent of the particular 1-parameter
family, although it depends on the type of collapsing matter.

\item In the region of large curvature before black hole formation,
the spacetime approaches a self-similar solution which is also
universal with respect to initial data, the {\em critical solution}. 

\end{itemize}

Critical phenomena were discovered by Choptuik 
\cite{Choptuik92} in numerical simulations of a spherical scalar
field. They have been found in numerous other numerical and
analytical studies in spherical symmetry, and a few in axisymmetry; in
particular critical phenomena have been seen in the collapse of
axisymmetric gravitational waves in vacuum \cite{AbrahamsEvans}. 
Scaling laws similar to the one for black hole mass 
have been discovered for black hole charge and
conjectured for black hole angular momentum.

It is still unclear how universal critical phenomena in collapse are
with respect to matter types and beyond spherical symmetry, in
particular for vacuum collapse. Progress is likely to be made over the
next few years with better numerical simulations.

Critical phenomena can usefully be described in dynamical systems
terms. A critical solution is then characterised as an attracting
fixed point within a surface that divides two basins of attraction, a
{\em critical surface} in phase space. Such a fixed point can be
either a stationary spacetime, or one that is scale-invariant and
self-similar. The latter is relevant for the (``type II'') critical
phenomena sketched above. We shall also see how the dynamical systems
approach establishes a connection with critical phenomena in
statistical mechanics.

Therefore, we could define the field of critical phenomena in
gravitational collapse as the study of the boundaries among the basins
of attraction of different end states of self-gravitating systems,
such as black hole formation or dispersion. In our view, the
main physical motivation for this study is that those critical solutions
which are self-similar provide a way of achieving arbitrarily large
spacetime curvature outside a black hole, and in the limit a naked
singularity, by fine-tuning generic initial data for generic matter
to the black hole threshold. Those solutions are therefore likely
to be important for quantum gravity and cosmic censorship.


\subsection{Plan of this review}

Faced with more material to review, we have attempted to make this
update shorter and more systematic than the original 1999 version. We
begin with the abstract theory in Sec.~\ref{section:theory}.
Since 1999, progress on the theory side
has mainly been made on the global spacetime structure of the critical
solution and cosmic censorship in the spherical scalar field model. We
have included this new material in an enlarged
Sec.~\ref{section:scalarfield} on the spherical scalar field, although
we hope it will turn out to be sufficiently generic to merit inclusion
in Sec.~\ref{section:theory}. Nonspherical perturbations of the
spherical scalar field are also discussed in
Sec.~\ref{section:scalarfield}.

In Sec.~\ref{section:spherical} we review the rich phenomenology that
has been found in many other systems restricted to spherical
symmetry. Numerical work in spherical symmetry has proliferated since
1999, but we have tried to keep this section as short as possible.
There has been less progress in going beyond spherical symmetry than
we anticipated in 1999, even though we continue to believe that
important results await there. What is known today is summarised in
Sec.~\ref{section:nonspherical}.

The reader unfamiliar with the topic is advised to begin
with either Secs.~\ref{section:universality}-\ref{section:scaling},
which give the key theory of universality, self-similarity and
scaling, or
Secs.~\ref{section:scalarfieldequations}-\ref{section:threshold},
which describe the classic example, the massless scalar field.

This review is limited to numerical and theoretical work on phenomena
at the threshold of black hole formation in 3+1-dimensional general
relativity. We report only briefly on work in higher and lower
spacetime dimensions and non-gravity systems that may be relevant as
toy models for general relativity. We exclude other work on 
self-similarity in general relativity and work on critical phenomena
in other areas of physics.

Other reviews on the subject are 
\cite{Horne_MOG},
\cite{Bizon},
\cite{Gundlach_Banach},
\cite{Gundlach_critreview1},
\cite{Choptuik_review},
\cite{Choptuik_review2},
\cite{BradyCai},
\cite{Gundlach_critreview2},
\cite{MartinGundlach2}.
The 2002 review \cite{gundlach_critreview3} by Gundlach gives
more detailed explanations on some of the basic aspects of the theory.


\section{Theory}
\label{section:theory}

In this Section we describe the basic theory underlying critical
collapse of the type that forms arbitrarily small black holes (later
called type II, see also Sec.~\ref{section:typeI}). We begin with the
mathematical origin of its three main characteristics, which were
already summarised in the introduction:

\begin{itemize}

\item universality with respect to initial data;

\item scale-invariance of the critical solution;

\item black hole mass scaling.

\end{itemize}


\subsection{Universality}
\label{section:universality}

Consider GR as an infinite-dimensional continuous dynamical
system. Points in the phase space are initial data
sets (3-metric, extrinsic curvature and suitable matter variables,
which together obey the Einstein constraints). We evolve with the Einstein
equations in a suitable gauge (see
Sec.~\ref{section:coordinates}). Solution curves of the dynamical
system are spacetimes obeying the Einstein-matter equations, sliced by
specific Cauchy surfaces of constant time $t$.

An isolated system in GR can end up in qualitatively different stable
end states. Two possibilities are the formation of a single
black hole in collapse, or complete dispersion of the mass-energy to
infinity. For a massless scalar field in spherical symmetry, these
are the only possible end states (see
Sec.~\ref{section:scalarfield}). Any point in phase space can be
classified as ending up in one or the other type of end state. The
entire phase space therefore splits into two halves, separated by a
``critical surface''.

A phase space trajectory that starts on a critical surface by
definition never leaves it. A critical surface is therefore a
dynamical system in its own right, with one dimension fewer than the
full system. If it has an attracting fixed point, such a point is
called a critical point. It is an attractor of codimension one in the
full system, and the critical surface is its attracting manifold. The
fact that the critical solution is an attractor of codimension one is
visible in its linear perturbations: it has an infinite number of
decaying perturbation modes spanning the tangent plane to the critical
surface, and a single growing mode not tangential to it.

\begin{figure}
\begin{center}
\includegraphics[width=10cm]{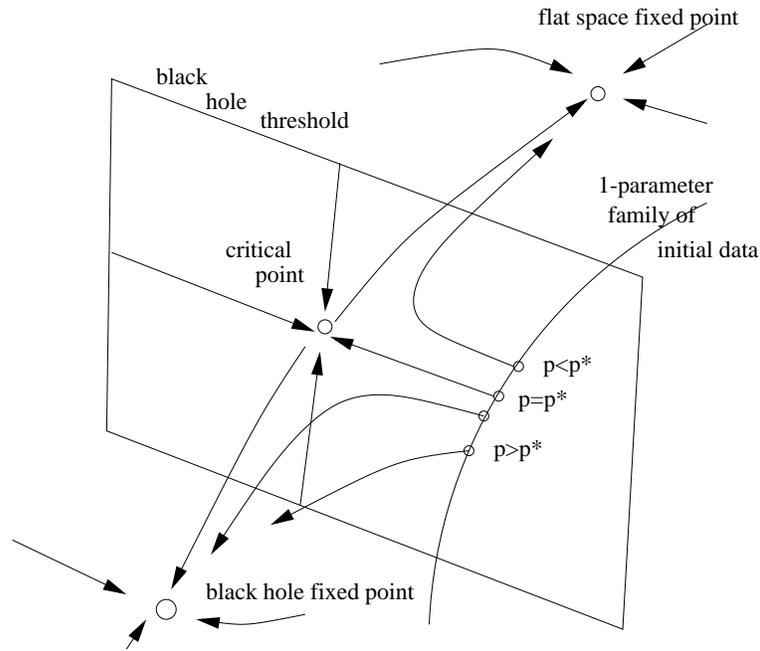}
\end{center}
\caption{
\label{fig:dynsim}
The phase space picture for the black hole threshold in the presence
of a critical point. Every point correspond to an initial data set,
that is, a 3-metric, extrinsic curvature, and matter fields. (In type
II critical collapse these are only up to scale). The arrow lines are
solution curves, corresponding to spacetimes, but the critical solution,
which is stationary (type I) or self-similar (type II) is represented
by a point. The line without an arrow is not a time evolution, but a
1-parameter family of initial data that crosses the black hole threshold
 at $p=p_*$. The 2-dimensional plane represents an
$(\infty-1)$-dimensional hypersurface, but the third dimension
represents really only one dimension.}
\end{figure}

\begin{figure}
\begin{center}
\includegraphics[width=15cm]{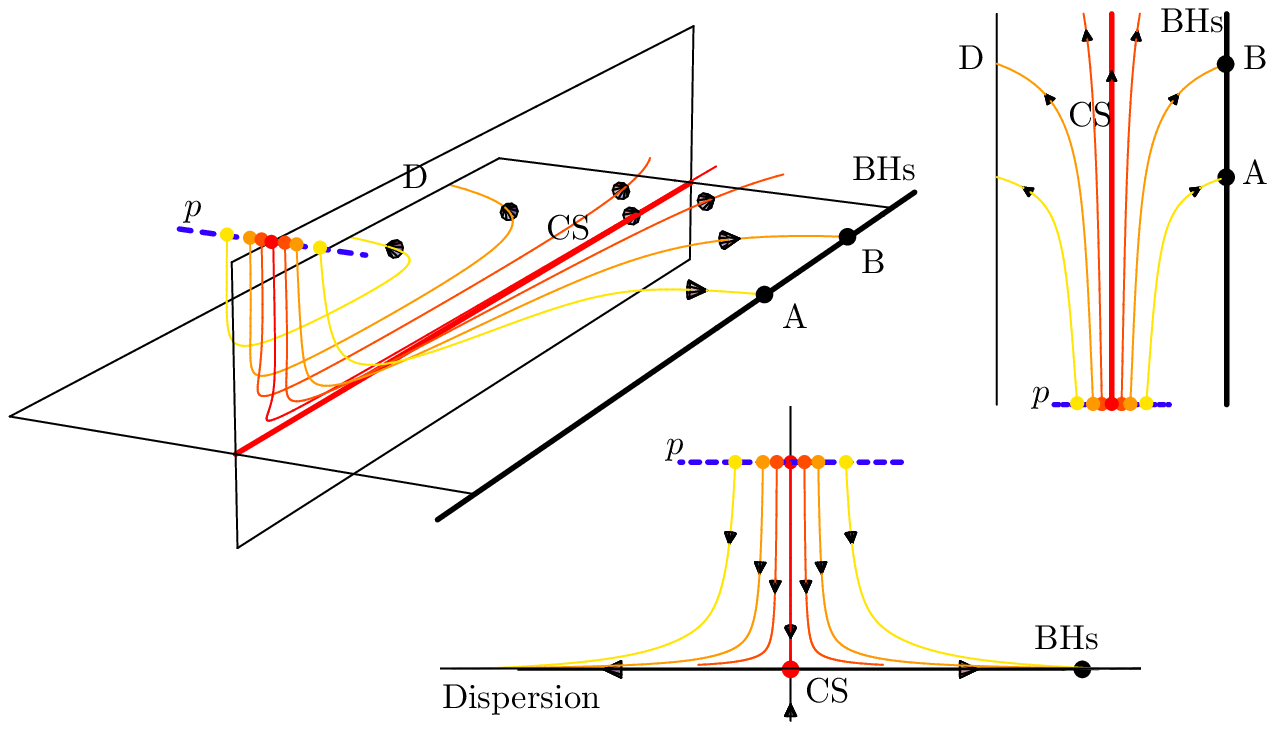}
\end{center}
\caption{\label{fig:phasespace3d} A different phase space picture,
specifically for type II critical collapse, and two 2-dimensional
projections of the same picture. In contrast with
Fig.~\ref{fig:dynsim}, one dimension of the two representing the
(infinitely many) decaying modes has been suppressed. The additional
axis now represents a global scale which was suppressed in
Fig.~\ref{fig:dynsim}, so that the scale-invariant critical solution
CS is now represented as a straight line (in red). Several members of
a family of initial conditions (in blue) are attracted by the critical
solution and then depart from it towards black hole formation (A or B)
or dispersion (D). Perfectly fine-tuned initial data asymptote to the
critical solution with decreasing scale. Initial conditions starting
closer to perfect fine tuning produce smaller black holes, such that
the parameter along the line of black hole endstates is $-\ln
M_{BH}$. Two 2-dimensional projections of the same picture are also
given. The horizontal projection of this figure is the same as the
vertical projection of Fig.~\ref{fig:dynsim}.}
\end{figure}

As illustrated in Figs.~\ref{fig:dynsim} and \ref{fig:phasespace3d},
any trajectory beginning near the critical surface, but not
necessarily near the critical point, moves almost parallel to the
critical surface towards the critical point. Near the critical point
the evolution slows down, and eventually moves away from the critical
point in the direction of the growing mode. This is the origin of
universality. All details of the initial data have been forgotten,
except for the distance from the black hole threshold. The closer the
initial phase point is to the critical surface, the more the solution
curve approaches the critical point, and the longer it will remain
close to it.  We should stress that this phase picture is extremely
simplified.  Some of the problems associated with this simplification
are discussed in Sec.~\ref{section:coordinates}.

\subsection{Self-similarity}

Fixed points of dynamical systems often have additional symmetries. In
the case of type II critical phenomena, the critical point is a
spacetime that is self-similar, or scale-invariant. 
These symmetries can be discrete or continuous. The critical
solution of a spherically symmetric perfect fluid (see
Sec.~\ref{section:perfectfluid}), has continuous self-similarity
(CSS). A CSS spacetime is one that admits a homothetic vector field
$\xi$, defined by \cite{CahillTaub}:
\begin{equation}
  \label{homothetic_metric}
  {\cal L}_\xi g_{ab} = 2 g_{ab}.
\end{equation}
In coordinates $x^\mu=(\tau, x^i)$ adapted to the symmetry, so that
\begin{equation}
  \label{xi_in_coordinates}
  \xi = - {\partial\over\partial\tau},
\end{equation}
the metric coefficients are of the form
\begin{equation}
  \label{CSS_coordinates}
  g_{\mu\nu}(\tau, x^i) = e^{-2\tau} \tilde g_{\mu\nu}(x^i),
\end{equation}
where the coordinate $\tau$ is the negative logarithm of a
spacetime scale, and the remaining three coordinates $x^i$ can be
thought of angles around the singular spacetime point $\tau=\infty$
(see Sec.~\ref{section:globalstructure}).

The critical solution of other systems, in particular the spherical
scalar field (see Sec.~\ref{section:scalarfield}) and axisymmetric
gravitational waves (see Sec.~\ref{section:brillwaves}), show 
discrete self-similarity (DSS). The simplest way of defining
DSS is in adapted coordinates, where
\begin{equation}
  \label{DSS_coordinates}
  g_{\mu\nu}(\tau, x^i) = e^{-2\tau} \tilde g_{\mu\nu}(\tau,x^i),
\end{equation}
such that $g_{\mu\nu}(\tau,x^i)$ is periodic in $\tau$ with period
$\Delta$. More formally, DSS can be defined as a discrete conformal
isometry~\cite{Gundlach_Chop1}.

Using the gauge freedom of general relativity, the lapse and shift in
the ADM formalism can be chosen (non-uniquely) so that the coordinates
become adapted coordinates if and when the solution becomes
self-similar (see
Sec.~\ref{section:coordinates}). $\tau$ is then both a time coordinate
(in the usual sense that surfaces of constant time are Cauchy
surfaces), and the logarithm of overall scale at constant $x^i$. The
minus sign in (\ref{xi_in_coordinates}) and hence
(\ref{CSS_coordinates}) and (\ref{DSS_coordinates}), is a convention
assuming that smaller scales are in the future. The time parameter
used in Fig.~\ref{fig:phasespace3d} is of this
type. 


\subsection{Mass scaling}
\label{section:scaling}

Let $Z$ stand for a set of scale-invariant variables of the problem,
such as $\tilde g_{\mu\nu}$ and suitably rescaled matter variables. If
the dynamics is scale-invariant (this is the case exactly for example
for the scalar field, and approximately for other systems, see
Sec.~\ref{section:universalityclasses}), then $Z(x)$ is an element of
the phase space factored by overall scale, and $Z(x,\tau)$ a
solution. Note that $Z(x)$ is an initial data set for GR only up to
scale. The overall scale is supplied by $\tau$.

For simplicity, assume that the critical solution is CSS. It can then
be written as $Z(x,\tau)=Z_*(x)$. Its linear perturbations can depend
on $\tau$ only exponentially. To linear order, the solution near the
critical point must be of the form
\begin{equation}
\label{perturbations}
  Z(x,\tau) \simeq Z_*(x) + \sum_{i=1}^\infty C_i(p) \, e^{\lambda_i
    \tau} Z_i(x).
\end{equation}
The perturbation amplitudes $C_i$ depend on the initial data, and
hence on $p$. As $Z_*$ is a critical solution, by definition there is
exactly one $\lambda_i$ with positive real part (in fact it is purely
real), say $\lambda_0$. As $\tau\to\infty$, all other perturbations
vanish. In the following we consider this limit, and retain only the
one growing perturbation. 

From our phase space picture, the evolution ends at the critical
solution for $p=p_*$, so we must
have $C_0(p_*)=0$. Linearising in $p$ around $p_*$, we obtain
\begin{equation}
  \label{echoing_region}
  \lim_{\tau\to\infty} Z(x,\tau) \simeq Z_*(x) + {dC_0\over dp} (p-p_*)
  e^{\lambda_0\tau} Z_0(x).
\end{equation}

For $p\ne p_*$, but close to it, the solution has the approximate form
(\ref{echoing_region}) over a range of $\tau$. Now we extract Cauchy
data at one particular $p$-dependent value of $\tau$ within that
range, namely $\tau_*$ defined by
\label{taustardef}
\begin{equation}
  {dC_0\over dp} (p-p_*) e^{-\lambda_0\tau_*} \equiv \epsilon,
\end{equation}
where $\epsilon$ is some constant $\ll 1$ such that at this $\tau$ the
linear approximation is still valid. At sufficiently large $\tau$, the
linear perturbation has grown so much that the linear approximation
breaks down, and for $C_0>0$ a black hole forms while for $C_0<0$ the
solution disperses. The crucial point is that we need not follow this
evolution in detail, nor does the precise value of $\epsilon$
matter. It is sufficient to note that the Cauchy data at $\tau=\tau_*$
are
\begin{equation}
\label{pdata}
  Z(x,\tau_*) \simeq Z_*(x) + \epsilon Z_0(x).
\end{equation}
Due to the funnelling effect
of the critical solution, the data
at $\tau_*$ is always the same, except for an overall scale, which is given
by $e^{-\tau_*}$. For example, the physical spacetime metric, with
dimension $(\hbox{length})^2$ is given by $g_{\mu\nu}=e^{-2\tau}
\tilde g_{\mu\nu}$, and similar scalings hold for the matter variables
according to their dimension.  In particular, as $e^{-\tau_*}$ is the
only scale in the initial data (\ref{pdata}), the mass of the final
black hole must be proportional to that scale. Therefore
\begin{equation}
\label{massscaling}
  M \propto e^{-\tau_*} \propto (p-p_*)^{1 / \lambda_0},
\end{equation}
and, comparing with Eq.~(\ref{eq1}), we have found the critical exponent
$\gamma = 1/\lambda_0$.

When the critical solution is DSS, a periodic or fine structure of
small amplitude is superimposed on this basic power law
\cite{Gundlach_Chop2, HodPiran_wiggle}:
\begin{equation}
  \ln M = \gamma \ln (p-p_*) + c + f\left(\gamma \ln (p-p_*) + c\right),
\end{equation}
where $f(z)$ has period $\Delta$ and is universal, and only $c$
depends on the initial data. As the critical solution is periodic in $\tau$
with period $\Delta$ the number $N$ of scaling ``echos'' is
approximated by
\begin{equation}
  N \simeq \Delta^{-1} \gamma \ln|p-p_*| + {\rm const.}
\end{equation}
Note that this holds for both supercritical and subcritical solutions.


\subsection{Type I}
\label{section:typeI}
 
In type I critical phenomena, the same phase space picture as in
Sec.~\ref{section:universality} applies, but the critical solution is
now stationary or time-periodic instead of self-similar or
scale-periodic. It also has a finite mass and can be thought of
as a metastable star. (Type I and II were so named after first and
second order phase transitions in statistical mechanics, in which the
order parameter is discontinuous and continuous, respectively.)
Universality in this context implies that the black hole mass near the
threshold is independent of the initial data, namely a certain
fraction of the mass of the stationary critical solution. The
dimensionful quantity that scales is not the black hole mass, but the
lifetime $t_p$ of the intermediate state where the solution is
approximated by the critical solution. This is clearly
\begin{equation}
\label{typeIscaling}
t_p = - {1\over \lambda_0} \ln|p-p_*| + {\rm const.}
\end{equation}

Type I critical phenomena occur when a mass scale in the field
equations becomes dynamically relevant. (This scale does not
necessarily set the mass of the critical solution absolutely: there
could be a family of critical solutions selected by the initial
conditions.) Conversely, as the type II power law is scale-invariant,
type II phenomena occur in situations where either the field equations
do not contain a scale, or this scale is dynamically irrelevant. Many
systems, such as the massive scalar field, show both type I and type
II critical phenomena, in different regions of the space of initial
data \cite{BradyChambersGoncalves}.


\subsection{Coordinate choices for the dynamical systems picture}
\label{section:coordinates}

The time evolution of Cauchy data in GR can only be considered as a
dynamical system if the ADM evolution equations are complemented by a
prescription for the lapse and shift. To realise the phase space
picture of Sec.~\ref{section:universality}, the critical solution
must be a fixed point or limit cycle. We have seen how coordinates
adapted to the self-similarity can be constructed, but is there a
prescription of the lapse and shift for {\em arbitrary} initial data,
such that, given initial data for the critical solution, the resulting
time evolution actively drives the metric to a form
(\ref{DSS_coordinates}) that explicitly displays the self-similarity?

Garfinkle and Gundlach \cite{GarfinkleGundlach} have suggested several
combinations of lapse and shift conditions that leave CSS spacetimes
invariant and turn the Choptuik DSS spacetime into a limit cycle.
(See \cite{GarfinkleMeyer,Garfinkle2} for
partial successes.) Among these, the combination of maximal slicing
with minimal strain shift has been suggested in a different context
but for related reasons \cite{SmarrYork}. Maximal slicing requires the
initial data slice to be maximal (${K_a}^a=0$), but other
prescriptions, such as freezing the trace of $K$ together with minimal
distortion, allow for an arbitrary initial slice with arbitrary
spatial coordinates.

All these coordinate conditions are elliptic equations that require
boundary conditions, and will turn CSS spacetimes into fixed points
(or DSS into limit cycles) only given correct boundary
conditions. Roughly speaking, these boundary conditions require a
guess of how far the slice is from the accumulation point $t=t_*$, and
answers to this problem only exist in spherical symmetry. Appropriate
boundary conditions are also needed if the dynamical system is
extended to include the lapse and shift as evolved variables, turning
the elliptic equations for the lapse and shift into hyperbolic or
parabolic equations.

Turning a CSS or stationary spacetime into a fixed point of the
dynamical system also requires an appropriate choice of the phase
space variables $Z(x^i)$. To capture CSS (or DSS)
solutions, one needs scale-invariant variables. Essentially, these can
be constructed by dimensional analysis. The coordinates $x^i$ and
$\tau$ are dimensionless, $le^{-\tau}$ has dimension length, and
$g_{\mu\nu}$ has dimension $l^2$. The scaling for the ADM and any
matter variables follows.

Even with a prescription for the lapse and shift in place, a given
spacetime does not correspond to a unique trajectory in phase
space. Rather, for each initial slice through the same spacetime one
obtains a different slicing of the entire spacetime. A possibility for
avoiding this ambiguity would be to restrict the phase space further,
for example by restricting possible data sets to maximal or constant
extrinsic curvature slices.

Another open problem is that in order to talk about attractors and
repellers on the phase space we need to define a norm on a suitable
function space which includes both asymptotically flat data and data for
the exact critical solution. The norm itself must favour the central region
and ignore what is further out and asymptotically flat if all black
holes of the same mass are to be considered as the same
endstate. 


\subsection{Approximate self-similarity and universality classes}
\label{section:universalityclasses}

The field equations for the massless scalar field coupled to the
Einstein equations are scale-free. Realistic matter models introduce
length scales, and the field equations then do not allow for
exactly self-similar solutions. They may however admit solutions which
are CSS or DSS asymptotically on small spacetime scales as the
dimensionful parameters become irrelevant, including type II critical
solutions \cite{Choptuik94,BradyChambersGoncalves,ChoptuikChmajBizon}.
This can be explored by a formal expansion in powers of the small
parameter $Le^{-\tau}$, where $L$ is a parameter with dimensions
length in the evolution equations. The zeroth order of the expansion
is the self-similar critical solution of the system with
$L=0$. A similar ansatz can be made for the linear perturbations of
the resulting background. The zeroth order of the background expansion
determines $\Delta$ exactly and independently of $L$, and the
zeroth order term of the linear perturbation expansion determines the
critical exponent $1/\lambda_0$ exactly, so that there is no need in
practice to calculate any higher orders in $L$ to make predictions for
type II critical phenomena where they occur. (With $L\ne
0$, the basin of attraction of the type II critical solution will
depend on $L$, and type I critical phenomena may also occur, see
Sec.~\ref{section:typeI}. A priori, there could also be more than one
type II critical solution for $L=0$, although this has not been
observed.)

This procedure has been carried out for the Einstein-Yang-Mills system
\cite{Gundlach_EYM} and for massless scalar electrodynamics
\cite{GundlachMartin}. Both systems have a single length scale $1/e$
(in geometric units $c=G=1$), where $e$ is the gauge coupling
constant. All values of $e$ can be said to form one universality class
of field equations~\cite{HaraKoikeAdachi} represented by $e=0$. This
notion of universality classes is fundamentally the same as in
statistical mechanics. Other examples include modifications to the
perfect fluid equation of state that do not affect the limit of high
density \cite{NeilsenChoptuik}. A simple example is 
that any scalar field potential $V(\phi)$ becomes dynamically
irrelevant compared to the kinetic energy $|\nabla\phi|^2$ in a
self-similar solution \cite{Choptuik94}, so that all scalar fields
with potentials are in the universality class of the free massless
scalar field.  Surprisingly, even two different models like the
$SU(2)$ Yang-Mills and $SU(2)$ Skyrme models in spherical symmetry are
members of the same universality class~\cite{BizonChmajTabor}.

If there are several scales $L_0$, $L_1$, $L_2$ etc.\ present in the
problem, a possible approach is to set the arbitrary scale in
(\ref{x_tau}) equal to one of them, say $L_0$, and define the
dimensionless constants $l_i=L_i/L_0$ from the others. The scope of
the universality classes depends on where the $l_i$ appear in the
field equations. If a particular $L_i$ appears in the field equations
only in positive integer powers, the corresponding $l_i$ appears only
multiplied by $e^{-\tau}$, and will be irrelevant in the scaling
limit. All values of this $l_i$ therefore belong to the same
universality class. From the example above, adding a quartic
self-interaction $\lambda\phi^4$ to the massive scalar field gives
rise to the dimensionless number $\lambda/m^2$ but its value is an
irrelevant (in the language of renormalisation group theory)
parameter.  

Contrary to the statement in~\cite{GundlachMartin}, we conjecture that
massive scalar electrodynamics, for any values of $e$ and $m$, is in
the universality class of the massless uncharged scalar field in a
region of phase space where type II critical phenomena occur. Examples
of dimensionless parameters which do change the universality class are
the $k$ of the perfect fluid, the $\kappa$ of the 2-dimensional sigma
model or, probably, a conformal coupling of the scalar field
\cite{Choptuik91} (the numerical evidence is weak but a dependence
should be expected).


\subsection{The analogy with critical phase transitions}
\label{section:analogy}


Some basic aspects of critical phenomena in gravitational collapse,
such as fine-tuning, universality, scale-invariant physics, and
critical exponents for dimensionful quantities, can also be identified
in critical phase transitions in statistical mechanics (see
\cite{Yeomans} for an introductory textbook).

From an abstract point of view, the objective of statistical mechanics
is to derive relations between macroscopic observables $A$ of the
system and macroscopic external forces $f$ acting on it, by
considering ensembles of microscopic states of the system. The
expectation values $\langle A\rangle$ can be generated as partial
derivatives of the partition function
\begin{equation}
Z(\mu, f)=\sum_{\rm microstates} \ e^{-  H({\rm
microstate}, \mu, f)}
\end{equation}
Here the $\mu$ are parameters of the Hamiltonian such as the strength
of intermolecular forces, and $f$ are macroscopic quantities which
are being controlled, such as the temperature or magnetic field.

Phase transitions in thermodynamics are thresholds in the space of
external forces $f$ at which the macroscopic observables $A$, or one
of their derivatives, change discontinuously. We consider two
examples: the liquid-gas transition in a fluid, and the ferromagnetic
phase transition.

The liquid-gas phase transition in a fluid occurs at the boiling curve
$p=p_b(T)$. In crossing this curve, the fluid density changes
discontinuously. However, with increasing temperature, the difference
between the liquid and gas density on the boiling curve decreases, and
at the the critical point $(p_*=p_b(T_*),T_*)$ it vanishes as a
non-integer power:
\begin{equation}
\rho_{\rm liquid}-\rho_{\rm gas}\sim (T_*-T)^\gamma.
\end{equation}
At the critical point an otherwise clear fluid becomes opaque, due to
density fluctuations appearing on all scales up to scales much larger
than the underlying atomic scale, and including the wavelength of
light. This indicates that the fluid near its critical point is
approximately scale-invariant (for some range of scales between the
size of molecules and the size of the container). 

In a ferromagnetic material at high temperatures, the magnetisation
${\bf m}$ of the material (alignment of atomic spins) is determined by
the external magnetic field ${\bf B}$. At low temperatures, the material
shows a spontaneous magnetisation even at zero external field. In the
absence of an external field this breaks rotational symmetry: the
system makes a random choice of direction. With increasing
temperature, the spontaneous magnetisation ${\bf m}$ decreases and
vanishes at the Curie temperature $T_*$ as
\begin{equation}
|{\bf m}|\sim (T_*-T)^\gamma.
\end{equation}
Again, the correlation length, or length scale of a typical
fluctuation, diverges at the critical point, indicating
scale-invariant physics. 

Quantities such as $|\bf m|$ or $\rho_{\rm liquid}-\rho_{\rm gas}$ are
called order parameters. In statistical mechanics, one distinguishes
between first-order phase transitions, where the order parameter
changes discontinuously, and second-order, or critical, ones, where it
goes to zero continuously. One should think of a critical phase
transition as the critical point where a line of first-order phase
transitions ends as the order parameter vanishes. This is already
clear in the fluid example. In the ferromagnet example, at first one
seems to have only the one parameter $T$ to adjust. But in the
presence of a very weak external field, the spontaneous magnetisation
aligns itself with the external field ${\bf B}$, while its strength is
to leading order independent of ${\bf B}$. The function ${\bf m}({\bf
B},T)$ therefore changes discontinuously at ${\bf B}=0$. The line
${\bf B}=0$ for $T<T_*$ is therefore a line of first order phase
transitions between directions (if we consider one spatial dimension
only, between ${\bf m}$ up and ${\bf m}$ down). This line ends at the
critical point $({\bf B}=0,T=T_*)$ where the order parameter $|{\bf
m}|$ vanishes. The critical value ${\bf B}=0$ of ${\bf B}$ is determined
by symmetry; by contrast $p_*$ depends on microscopic properties of
the material.

We have already stated that a critical phase transition involves
scale-invariant physics. In particular, the atomic scale, and any
dimensionful parameters associated with that scale, must become
irrelevant at the critical point. This is taken as the starting point
for obtaining properties of the system at the critical point.

One first defines a semi-group acting on micro-states: the
renormalisation group. Its action is to group together a small number
of adjacent particles as a single particle of a fictitious new system
by using some averaging procedure. This can also be done in a more
abstract way in Fourier space. One then defines a dual action of the
renormalisation group on the space of Hamiltonians by demanding that
the partition function is invariant under the renormalisation group
action:
\begin{equation}
\sum_{\rm microstates} e^{-H}=\sum_{\rm microstates'} e^{-H'}. 
\end{equation}
The renormalised Hamiltonian is in general more complicated than the
original one, but it can be approximated by the same Hamiltonian with
new values of the parameters $\mu$ and external forces $f$. (At this
stage it is common to drop the distinction between $\mu$ and $f$, as
the new $\mu'$ and $f'$ depend on both $\mu$ and $f$.) Fixed points of
the renormalisation group correspond to Hamiltonians with the
parameters at their critical values. The critical values
of many of these parameters will be $0$ or $\infty$, meaning that
the dimensionful parameters they were originally associated with
are irrelevant. Because a fixed point of the renormalisation group can
not have a preferred length scale, the only parameters that can have
nontrivial values are dimensionless.

The behaviour of thermodynamical quantities at the critical point is in
general not trivial to calculate. But the action of the
renormalisation group on length scales is given by its definition. The
blowup of the correlation length $\xi$ at the critical point is
therefore the easiest critical exponent to calculate. The same is true
for the black hole mass, which is just a length scale. We can
immediately reinterpret the mathematics of Section
\ref{section:scaling} as a calculation of the critical exponent for
$\xi$, by substituting the correlation length $\xi$ for the black hole
mass $M$, $T_*-T$ for $p-p_*$, and taking into account that the
$\tau$-evolution in critical collapse is towards smaller scales, while
the renormalisation group flow goes towards larger scales: $\xi$
therefore diverges at the critical point, while $M$ vanishes.

In type II critical phenomena in gravitational collapse, we should
think of the black hole mass as being controlled by the functions $P$
and $Q$ on phase space defined by Eq.~(\ref{PQ}). Clearly, $P$ is the
equivalent of the reduced temperature $T-T_*$. Gundlach
\cite{Gundlach_scalingfunctions} has suggested that the angular
momentum of the initial data can play the role of ${\bf B}$, and the
final black hole angular momentum the role of ${\bf m}$. Like the
magnetic field, angular momentum is a vector, with a critical value
that must be zero because all other values break rotational symmetry.


\section{The scalar field}
\label{section:scalarfield}

Critical phenomena in gravitational collapse were first discovered by
Choptuik \cite{Choptuik91, Choptuik92, Choptuik94} in the model of a
spherically symmetric, massless scalar field $\phi$ minimally coupled to
general relativity. The scalar field matter is both simple, and acts
as a toy model in spherical symmetry for the effects of gravitational
radiation. 
Given that it is still the best-studied model
in spherical symmetry, we review it here as a case study. For other
numerical work on this model, see \cite{GPP2, Garfinkle,
HamadeStewart, FrolovPen, Puerrer}. Important analytical studies of
gravitational collapse in this model have been carried out by
Christodoulou \cite{Christodoulou0, Christodoulou0a, Christodoulou1,
Christodoulou2, Christodoulou3, Christodoulou4, Christodoulou5}.

We first review the field equations and Choptuik's observations at
the black hole threshold, mainly as a concrete example for the general
ideas discussed above. We then summarise more recent work on the
global structure of Choptuik's critical solution, which throws an
interesting light on cosmic censorship. In particular, the exact
critical solution contains a curvature singularity that is locally and
globally naked, and any critical solution obtained in the limit of
perfect fine-tuning of asymptotically flat initial data is at least
locally naked. By perturbing around spherical symmetry, the
stability of the Choptuik solution in the full phase space can be
investigated, and the scaling of black hole angular momentum can be
predicted. By embedding the real scalar field in scalar
electrodynamics and perturbing around the Choptuik solution, the
scaling of black hole charge can be predicted.


\subsection{Field equations in spherical symmetry}
\label{section:scalarfieldequations}

The Einstein equations are
\begin{equation}
  \label{scalar_stress_energy}
  G_{ab} = 8 \pi \left(\nabla_a \phi \nabla_b \phi - {1\over 2} g_{ab}
  \nabla_c \phi \nabla^c \phi\right),
\end{equation}
and the matter equation is
\begin{equation}
  \nabla_a \nabla^a \phi = 0.
\end{equation}
Note that the matter equation of motion is contained within the
contracted Bianchi identities. Choptuik chose Schwarzschild-like
coordinates 
\begin{equation}
  \label{tr_metric}
  ds^2 = - \alpha^2(r, t) \, dt^2 + a^2(r, t) \, dr^2 + r^2 \, d\Omega^2,
\end{equation}
where $d\Omega^2 = d\theta^2 + \sin^2\theta \, d\varphi^2$ is the
metric on the unit 2-sphere. This choice of coordinates is defined by
the radius $r$ giving the surface area of 2-spheres as $4\pi r^2$, and by
$t$ being orthogonal to $r$ (polar-radial coordinates). One more
condition is required to fix the coordinate completely. Choptuik chose
$\alpha=1$ at $r=0$, so that $t$ is the proper time of the central
observer.

In the auxiliary variables
\begin{equation}
  \Phi = \phi_{, r}, \qquad \Pi={a\over\alpha} \phi_{, t},
\end{equation}
the wave equation becomes a first-order system,
\begin{equation}
  \label{wave}
  \begin{array}{l}
    \displaystyle \Phi_{, t} = \left({\alpha\over a}\Pi\right)_{, r},
    \\ \\
    \displaystyle \Pi_{, t} = {1\over r^2} \left(r^2{\alpha\over a}
    \Phi\right)_{, r}.
  \end{array}
\end{equation}
In spherical symmetry there are four algebraically independent
components of the Einstein equations. Of these, one is a linear
combination of derivatives of the other and can be disregarded. The
other three contain only first derivatives of the metric, namely
$a_{, t}$, $a_{, r}$ and $\alpha_{, r}$, and are
\begin{eqnarray}
  \label{da_dr}
  {a_{, r}\over a}  + {a^2 -1 \over 2r} 
  &=& 2\pi r (\Pi^2 + \Phi^2), \\
  \label{dalpha_dr}
  {\alpha_{, r}\over \alpha}  - {a_{, r}\over a}  - {a^2 -1 \over r}
  &=& 0, \\
  \label{da_dt}
  {a_{, t}\over \alpha}  &=& 4\pi r \Phi \Pi.
\end{eqnarray}
Because of spherical symmetry, the only dynamics is in the scalar
field equations (\ref{wave}). The metric can be found by integrating
the ODEs (\ref{da_dr}) and (\ref{dalpha_dr}) for $a$ and $\alpha$ at
any fixed $t$, given $\phi$ and $\Pi$. Eq.~(\ref{da_dt}) can be
ignored in this ``fully constrained'' evolution scheme.


\subsection{The black hole threshold}
\label{section:threshold}

The free data for the system are the two functions $\Pi(r,0)$ and
$\Phi(r,0)$. Choptuik investigated several one-parameter families of
such data by evolving the data for many different values of the
parameter. Simple examples of such families are $\Pi(r,0)=0$ and a
Gaussian for $\Phi(r,0)$,
with the parameter $p$ taken to be either the
amplitude of the Gaussian, with the width and centre fixed, or the
width, with position and amplitude fixed, or the position, with width
and amplitude fixed. For sufficiently small amplitude (or the peak
sufficiently wide), the scalar field will disperse, and for
sufficiently large amplitude it will form a black hole. 

Generic one-parameter families behave in this way, but this is
difficult to prove in generality. Christodoulou showed for the
spherically symmetric scalar field system that data sufficiently weak
in a well-defined way evolve to a Minkowski-like
spacetime~\cite{Christodoulou0a, Christodoulou3}, and that a class of
sufficiently strong data forms a black hole~\cite{Christodoulou2}. 

Choptuik found that in all 1-parameter families of initial data he
investigated he could make arbitrarily small black holes by
fine-tuning the parameter $p$ close to the black hole threshold. An
important fact is that there is nothing visibly special to the black
hole threshold. One cannot tell that one given data set will form a
black hole and another one infinitesimally close will not, short of
evolving both for a sufficiently long time. 

As $p\to p_*$ along the family, the spacetime varies on ever smaller
scales. Choptuik developed numerical techniques that recursively
refine the numerical grid in spacetime regions where details arise on
scales too small to be resolved properly. In the end, he could
determine $p_*$ up to a relative precision of $10^{-15}$, and make
black holes as small as $10^{-6}$ times the ADM mass of the
spacetime. The power-law scaling (\ref{massscaling}) was obeyed from
those smallest masses up to black hole masses of, for some families,
$0.9$ of the ADM mass, that is, over six orders of
magnitude~\cite{Choptuik94}. There were no families of initial data
which did not show the universal critical solution and critical
exponent. Choptuik therefore conjectured that $\gamma$ is the same for
all one-parameter families of smooth, asymptotically flat initial data
that depend smoothly on the parameter, and that the approximate
scaling law holds ever better for arbitrarily small $p-p_*$.

It is an empirical fact that typical one-parameter families cross the
threshold only once, so that there is every indication that it is a
smooth submanifold, as we assumed in the phase space picture. Taking
into account the discussion of mass scaling above, we can formally
write the black hole mass as a functional of the initial data
$z=\left(\phi(r,0),\Pi(r,0)\right)$ {\em exactly} as
\begin{equation}
\label{PQ}
  M[z] = Q[z] H(P[z])\,(P[z])^\gamma, 
\end{equation}
where $P$ and $Q$ are {\em smooth} functions on phase space and $H$ is
the Heaviside function. ($Q$ could be absorbed into $P$).

In hindsight, polar-radial gauge is well-adapted to
self-similarity. In this gauge, discrete self-similarity corresponds
to
\begin{equation}
  \label{tr_scaling}
  Z(r, t) = Z\left(e^{n\Delta}r, e^{n\Delta}t\right)
\end{equation}
for any integer $n$, where $Z$ stands for any one of the
dimensionless quantities $a$, $\alpha$ or
$\phi$ (and therefore also for $r\Pi$ or $r\Phi$). With 
\begin{equation}
  \label{x_tau}
  x = -{r \over t-t_*}, \quad \tau = - \ln\left(-{t-t_*\over
  L}\right), \quad t<t_*,
\end{equation}
discrete self-similarity is
\begin{equation}
  Z(x,\tau+\Delta) = Z(x,\tau).
\end{equation}
The dimensionful constants $t_*$ and $L$ depend on the particular
one-parameter family of solutions, but the dimensionless critical
fields $a_*$, $\alpha_*$ and $\phi_*$, and in particular their
dimensionless period $\Delta$, are universal. Empirically,
$\Delta\simeq 3.44$ for the scalar field in numerical time evolutions,
and $\Delta=3.445452402(3)$ from a numerical construction of the
critical solution based on exact self-similarity and analyticity
\cite{critcont}.


\subsection{Global structure of the critical solution}
\label{section:globalstructure}

In adapted coordinates, the metric of the critical spacetime is of the
form $e^{-2\tau}$ times a regular metric. From this general form
alone, one can conclude that $\tau=\infty$ is a curvature singularity,
where Riemann and Ricci invariants blow up like $e^{4\tau}$ (unless
the spacetime is flat), and which is at finite proper time from
regular points in its past.  The Weyl tensor with index position
${C^a}_{bcd}$ is conformally invariant, so that components with this
index position remain finite as $\tau\to\infty$. This type of
singularity is called ``conformally compactifiable''~\cite{Tod_pc} or
``isotropic''~\cite{Goodeetal}. For a classification of all possible
{\em global} structures of spherically symmetric self-similar spacetimes see
\cite{GundlachMartinCSS}.

The global structure of the scalar field critical solution was
determined accurately in \cite{critcont} by assuming analyticity at
the centre of spherical symmetry and at the past light cone of the
singularity (the self-similarity horizon, or SSH). The critical
solution is then analytic up to the future lightcone of the
singularity (the Cauchy horizon, or CH). Global adapted coordinates
$x$ and $\tau$ can be chosen so that the regular centre $r=0$, the SSH
and the CH are all lines of constant $x$, and surfaces of constant
$\tau$ are never tangent to $x$ lines. (A global $\tau$ is no longer a global
time coordinate). This is illustrated in
Fig.~\ref{figure:redshiftmatched}.

Approaching the Cauchy horizon, the scalar field oscillates an
infinite number of times but with the amplitude of the oscillations
decaying to zero. The scalar
field in regular adapted coordinates $(x,\tau)$ is of the form
\begin{equation}
\label{phicritcont}
\phi(x,\tau)\simeq F_\mathrm{reg}(\tau)
+|x|^\epsilon F_\mathrm{sing}\Bigl[
\tau+H(\tau)+K\ln|x|\Bigr],
\end{equation}
where $F_\mathrm{reg}(\tau)$, $F_\mathrm{sing}(\tau)$ and $H(\tau)$
are periodic with period $\Delta$, and the SSH is at $x=0$. These
functions have been computed numerically to high accuracy, together
with the constants $K$ and $\epsilon$. The scalar
field itself is smooth with respect to $\tau$, and as $\epsilon>0$, it
is continuous but not
differentiable with respect to $x$ on the CH itself. The same is true
for the metric and the curvature. Surprisingly, the ratio $m/r$
of the Hawking mass over the area radius on the Cauchy horizon is of
order $10^{-6}$ but not zero (the value is known to eight significant
figures). 

As the CH itself is regular with smooth null data except for the
singular point at its base, it is not intuitively clear why the
continuation is not unique. A partial explanation is given in
\cite{critcont}, where all {\em DSS} continuations are considered.
Within a DSS ansatz, the solution just to the future of the CH has the
same form as (\ref{phicritcont}).  $F_\mathrm{reg}(\tau)$ is the same
on both sides, but $F_\mathrm{sing}(\tau)$ can be chosen freely on the
future side of the CH. Within the restriction to DSS this function can
be taken to parameterise the information that comes out of the naked
singularity.

There is precisely one choice of $F_\mathrm{sing}(\tau)$ on the future
side that gives a regular centre to the future of the CH, with the
exception of the naked singularity itself, which is then a point. This
continuation was calculated numerically, and is almost but not quite
Minkowski in the sense that $m/r$ remains small everywhere to the
future of the SSH. 

\begin{figure}
\begin{center}
\includegraphics[width=7cm]{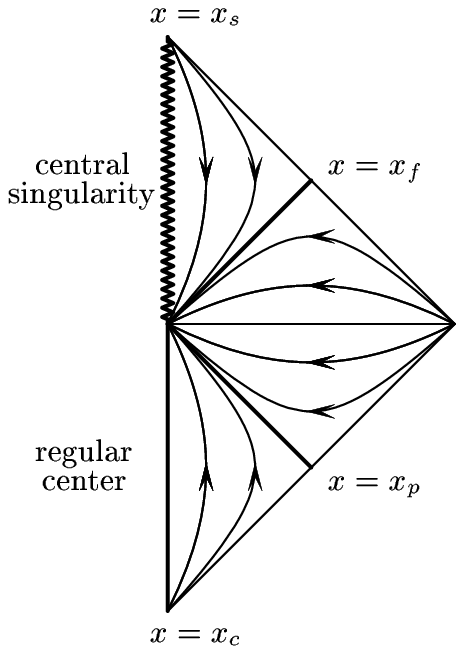}
\end{center}
\caption{
\label{figure:critcont}
The spacetime diagram
of all generic {\em DSS} continuations of the scalar field critical
solution, from~\cite{critcont}. The naked singularity is timelike,
central, strong, and has negative mass. There is also a unique
continuation where the singularity is replaced by a regular centre
except at the spacetime point at the base of the CH, which is still a
strong curvature singularity. No other spacetime diagram is possible
if the continuation is DSS. The lines with arrows are lines of
constant adapted coordinate $x$, with the arrow indicating the
direction of $\partial/\partial\tau$ towards larger curvature.}
\end{figure}

All other DSS continuations have a naked, timelike
central curvature singularity with negative mass. More exotic
continuations including further Cauchy horizons would be allowed
kinematically \cite{CarrGundlach} but are not achieved dynamically
if we assume that the continuation is DSS. The spacetime diagram of the
generic DSS continued solution is given in Fig.~\ref{figure:critcont}.

\subsection{Near-critical spacetimes and naked singularities}

Choptuik's results have an obvious bearing on the issue of cosmic
censorship. (See~\cite{Wald_censorship} for a general review of cosmic
censorship.) Roughly speaking, fine-tuning to the black hole threshold
provides a set of data which is codimension one in the space of
generic, smooth, asymptotically flat initial data, and whose time
evolution contains at least the point singularity of the critical
solution. The cosmic censorship hypothesis must therefore be
formulated as ``{\em generic} smooth initial data for reasonable
matter do not form naked singularities''. Here we look at the relation
between fine-tuning and naked singularities in more detail. 

Christodoulou \cite{Christodoulou5} proves rigorously that naked singularity
formation is not generic, but in a rather larger function space,
functions of bounded variation, than one would naturally consider. In
particular, the instability of the naked singularity found by
Christodoulou is not differentiable on the past light cone. This
is unnatural in the context of critical collapse, where the naked
singularity can arise from generic (up to fine-tuning) smooth initial
data, and the intersection of the past light cone of the singularity
with the initial data surface is as smooth as the initial data
elsewhere. It is therefore not clear how this theorem relates to the
numerical and analytical results strongly indicating that naked
singularities are codimension-1 generic within the space of
smooth initial data.

\begin{figure}
\begin{center}
\includegraphics[width=12cm]{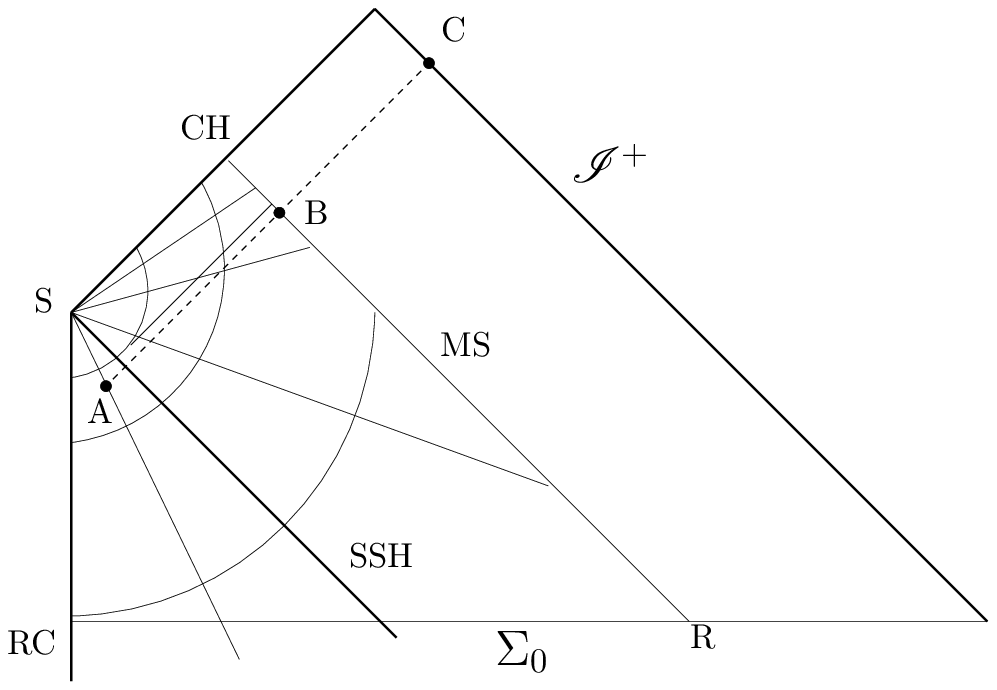}
\end{center}
\caption{
\label{figure:redshiftmatched}
Conformal diagram of the critical solution matched to an
asymptotically flat one. RC=regular centre, S=Singularity,
SSH=self-similarity horizon. Curved lines are lines of constant
coordinate $\tau$, while converging straight lines are lines of
constant coordinate $x$. Let the
initial data on the Cauchy surface CS be those for the exact critical
solution out to the 2-sphere R, and let these data be smoothly
extended to some data that are asymptotically flat, so that the future
null infinity \scri exists. To the past of the matching surface MS the solution
coincides with the critical solution. The spacetime cannot be uniquely
continued beyond the Cauchy horizon CH. The redshift from point A to point B is
finite by self-similarity, and the redshift from B to C is finite by
asymptotic flatness.}
\end{figure}

First, consider the exact critical solution. The lapse $\alpha$
defined by (\ref{tr_metric}) is bounded above and below in the
critical solution. Therefore the redshift measured between constant
$r$ observers located at any two points on an outgoing radial null
geodesic in the critical spacetime to the past of the CH is bounded
above and below. Within the exact critical solution, a point with
arbitrarily high curvature can therefore be observed from a point with
arbitrarily low curvature. Next consider a spacetime where the critical
solution in a central region is smoothly matched to an asymptotically
flat outer region such that the resulting asymptotically
flat spacetime contains a part of the critical solution that includes the
singularity and a part of the CH. In this spacetime, a point of
arbitrarily large curvature can be seen from \scri with finite
redshift. This is illustrated in Fig.~\ref{figure:redshiftmatched}.

Now consider the evolution of asymptotically flat initial data that
have been fine-tuned to the black hole threshold. The global structure
of such spacetimes has been investigated numerically in
\cite{HamadeStewart,FrolovPen,Garfinkle,Puerrer}. Empirically, these
spacetimes can be approximated near the singularity by 
(\ref{perturbations}). Almost all perturbations decay as the
singularity is approached and the approximation becomes better, until
the one growing perturbation (which by the assumption of fine-tuning
starts out small) becomes significant. A maximal value of the
curvature is then reached which is still visible from \scri and which
scales as \cite{GarfinkleDuncan}
\begin{equation}
R_\mathrm{max}\propto(p-p_*)^{2\gamma}.
\end{equation}
Finally, consider the limit of perfect fine-tuning. The growing mode
is then absent, and all other modes decay as the naked singularity is
approached. Note that (\ref{perturbations}) suggests this is true
regardless of the direction (future, past, or spacelike, depending on
the value of $x$) in which the singularity is
approached. This seems to be in conflict with causality: the issue is
sensitively connected to the completeness of the modes $Z_i$ and to
the stability of the Cauchy horizon and requires more investigation.

\begin{figure}
\begin{center}
\includegraphics[width=12cm]{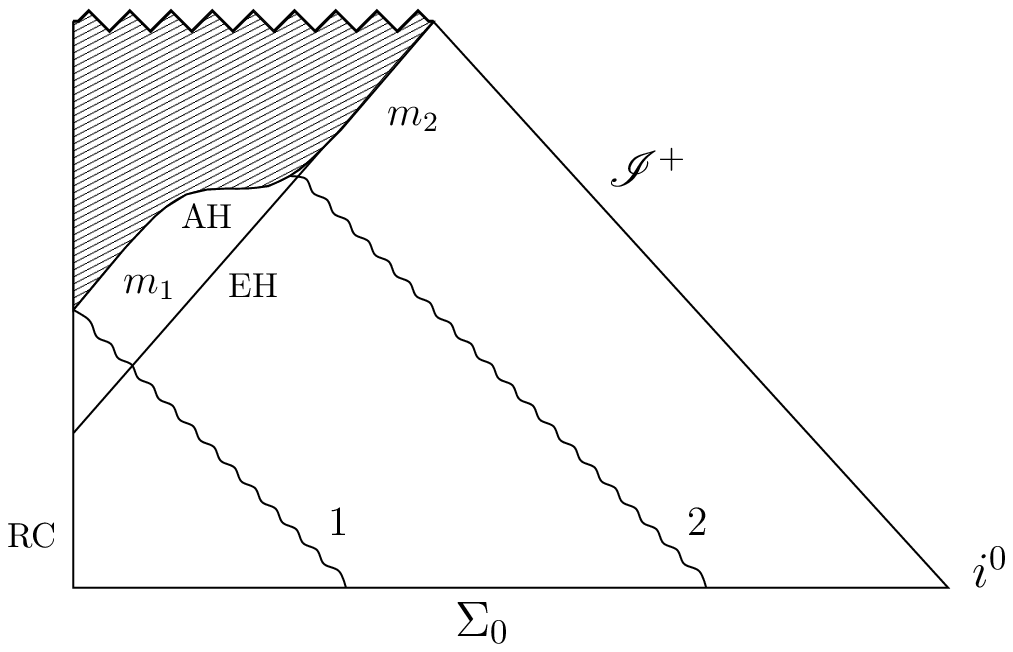}
\end{center}
\caption{\label{figure:smallBHlargeBH}
The final event horizon of a
black hole is only known when the infall of matter has
stopped. Radiation at 1 collapses to form a small black hole which
settles down, but later more radiation at 2 falls in to give rise to a
larger final mass. Fine-tuning of a parameter may result in $m_1\sim
(p-p_*)^\gamma$, but the final mass $m_2$ would be approximately
independent of $p$. 
}
\end{figure}

A complication in the supercritical case has been pointed out in
\cite{Puerrer}. In the literature on supercritical evolutions, what
is quoted as the black hole mass is in fact the mass of the first
apparent horizon (AH) that appears in the time slicing used by the
code (spacelike or null). The black hole mass can and generically will
be larger than the AH mass when it is first measured because of matter
falling in later, and the region of maximal curvature may well be
inside the event horizon, and hidden from observers at \scri.
(See Fig.~\ref{figure:smallBHlargeBH} for an illustration).
The true black hole mass can only be measured at \scri, where it is
defined to be the limit of the Bondi mass $m_\mathrm{B}$ as the Bondi
time $u_\mathrm{B}\to\infty$. This was implemented in
\cite{Puerrer}. Only one family of initial data was investigated, but
in this family it was found that $m_\mathrm{B}$ converges to $10^{-4}$
of the initial Bondi mass in the fine tuning limit. More numerical
evidence would be helpful, but the result is plausible. As the
underlying physics is perfectly scale-invariant in the massless scalar
field model, the minimum mass must be determined by the family of
initial data through the infall of matter into the black hole. 
Simulations of critical collapse of a perfect fluid in a cosmological
context show a similar lower bound \cite{HawkeStewart} due to matter
falling back after shock formation, but this may not be true for all
initial data \cite{MuscoMiller}.

\subsection{Electric charge}
\label{section:scalarcharge}

Given the scaling power law for the black hole mass in critical
collapse, one would like to know what happens if one takes a generic
1-parameter family of initial data with both electric charge and
angular momentum (for suitable matter), and fine-tunes the parameter
$p$ to the black hole threshold. A simple model for charged matter is a
complex scalar field coupled to electromagnetism with the substitution
$\nabla_a\to\nabla_a+ieA_a$, or scalar electrodynamics. (Note that in
geometric units, black hole charge $Q$ has dimension of length, but
the charge parameter $e$ has dimension 1/length.)

Gundlach and Mart\'\i n-Garc\'\i a \cite{GundlachMartin} have studied
scalar massless electrodynamics in spherical symmetry
perturbatively. Clearly, the real scalar field critical solution of
Choptuik is a solution of this system too. In fact, it remains
a critical solution within massless scalar electrodynamics in the
sense that it still has only one growing perturbation mode within the
enlarged solution space. Some of its perturbations carry electric
charge, but as they are all decaying, electric charge is a subdominant
effect. The charge of the black hole in the critical limit is
dominated by the most slowly decaying of the charged modes. From this
analysis, a universal power-law scaling of the black hole charge
\begin{equation}
Q\sim (p-p_*)^\delta
\end{equation}
was predicted. The predicted value $\delta\simeq 0.88$ of the critical
exponent (in scalar electrodynamics) was subsequently verified in
collapse simulations by Hod and Piran \cite{HodPiran_charge} 
and later again by Petryk \cite{Petryk}. (The
mass scales with $\gamma\simeq 0.37$ as for the uncharged scalar
field.) No other type of criticality can be found in the phase
space of this system as dispersion and black holes are the only
possible end states, though black holes with $|Q|\lesssim M $ can
be formed \cite{Petryk}.

General considerations similar to those in
Sec.~\ref{section:universalityclasses} led Gundlach and Mart\'\i
n-Garc\'\i a to the general prediction that the two critical exponents
are always related, for any matter model, by the inequality
\begin{equation}
\delta\ge2\gamma
\end{equation}
(with the equality holding if the critical solution is charged),
so that black hole charge can always be treated perturbatively at the
black hole threshold. This has not yet been verified in any other
matter model.


\subsection{Self-interaction potential}
\label{section:self-interaction}

An example of the richer phenomenology in the presence of a scale in
the field equations is the spherical {\em massive} scalar field with a
potential $m^2\phi^2$~\cite{BradyChambersGoncalves} coupled to
gravity: In one region of phase space, with characteristic scales
smaller than $1/m$, the black hole threshold is dominated by the
Choptuik solution and type II critical phenomena occur. In another it
is dominated by metastable oscillating boson stars (whose mass is of
order $1/m$ in geometric units) and type I critical phenomena
occur. (For the real scalar field, the type I critical solution
is an (unstable) oscillating boson star \cite{SeidelSuen} while for
the complex scalar field it can be a static (unstable) boson star
\cite{HawleyChoptuik}.)

When the scalar field with a potential is coupled 
to electromagnetism, type II criticality is still
controlled by a solution which asymptotically resembles the uncharged
Choptuik spacetime, but type I criticality is now controlled by
charged boson stars \cite{Petryk}. There are indications that
subcritical type I evolutions lead to slow, large amplitude
oscillations of stable boson stars \cite{Lai, LaiChoptuik, Petryk}
and not to dispersion to infinity, as had been conjectured in
\cite{HawleyChoptuik}. 
Another interesting extension is the study of the dynamics of a real
scalar field with a symmetric double-well potential, in which the
system displays type I criticality between the two possible vacua
\cite{HondaChoptuik}. 


\subsection{Nonspherical perturbations: stability and angular
  momentum}
\label{section:scalarnonspherical}

Critical collapse is really relevant for cosmic censorship only if it
is not restricted to spherical symmetry. Mart\'{\i}n-Garc\'{\i}a and
Gundlach \cite{MartinGundlach} have analysed all nonspherical
perturbations of the scalar field critical solution by solving a
linear eigenvalue problem with an ansatz of regularity at the centre
and the SSH. They find that the only growing mode is the known
spherical one, while all other spherical modes and all non-spherical
modes decay. This strongly suggests that the critical solution is an
attractor of codimension one not only in the space of spherically
symmetric data but (modulo linearisation stability) of all data in a
finite neighbourhood of spherical symmetry.

More recently, Choptuik and collaborators \cite{Choptuik_axisymmetry}
have carried out axisymmetric time evolutions for the massless scalar
field using adaptive mesh refinement. They find that in the limit of
fine-tuning generic axisymmetric initial data the spherically
symmetric critical solution is approached at first but then deviates
from spherical symmetry and eventually develops two centres, each of
which approaches the critical solution and bifurcates again in a
universal way. This suggests that the critical solution has
non-spherical growing perturbation modes, possibly a single $l=2$ even
parity mode (in axisymmetry, only $m=0$ is allowed). There appears to be a conflict between the time evolution
results \cite{Choptuik_axisymmetry} and the perturbative results
\cite{MartinGundlach}, which needs to be resolved by more work (see
Sec.~\ref{section:brillwaves}).

Perturbing the scalar field around spherical symmetry, angular
momentum comes in to second order in perturbation theory. All angular
momentum perturbations were found to decay, and a critical exponent
$\mu\simeq 0.76$ for the angular momentum was derived for the massless
scalar field in \cite{GarfinkleGundlachMartin}. This prediction has
not yet been tested in nonlinear collapse simulations. 


\section{More spherical symmetry}
\label{section:spherical}

The pioneering work of Choptuik on the spherical massless scalar field
has been followed by a plethora of further investigations. These could
be organised under many different criteria. We have chosen the
following rough categories:

\begin{itemize}

\item Systems in which the field equations, {\em when reduced to
  spherical symmetry}, form a single wave-like equation, typically
  with explicit $r$-dependence in its coefficients. This
  includes Yang-Mills fields, sigma models, vector and spinor fields,
  scalar fields in 2+1 or in 4+1 and more spacetime dimensions,
  and scalar fields in a semi-classical approximation to quantum
  gravity. 

\item Perfect fluid matter, either in an asymptotically flat or a
  cosmological context. The linearised Euler equations are in fact
  wave-like, but the full non-linear equations admit shock heating and
  are therefore not even time-reversal symmetric. 

\item Collisionless matter described by the Vlasov equation 
  is a partial differential equation on particle phase space as
  well as spacetime. Therefore
  even in spherical symmetry, the matter equation is a partial
  differential equation in four dimensions (rather than two). Intuitively
  speaking, there are infinitely more matter degrees of freedom than
  in the scalar field or in non-spherical vacuum gravity.

\item Spherically symmetric nonlinear wave equations on (3+1) Minkowski
  spacetime, and other nonlinear partial differential equations which
  show a transition between singularity formation and dispersal. 

\end{itemize}

Some of these examples were constructed because they may have intrinsic
physical relevance (semiclassical gravity, primordial black holes),
others as toy models for 3+1-dimensional gravity, and others mostly 
out of a purely mathematical interest. Table~\ref{table:mattermodels}
gives an overview of these models.


\subsection{Matter obeying wave equations}

\subsubsection{2d nonlinear $\sigma$ model}
We have already discussed in subsection \ref{section:self-interaction}
the effects of adding a potential to the evolution of the scalar field.
An alternative generalization is a modification of the kinetic term in
the Lagrangian, with the general form
\begin{equation}
\frac{1}{2} g^{\mu\nu} G_{IJ}(\phi^K)\nabla_\mu\phi^I \nabla_\nu \phi^J
\end{equation}
for an N-dimensional vector field $\phi^I(x)$ with $I=1..N$, and where
$G_{IJ}(\phi^K)$ is a fixed nonlinear function acting as
a metric on the so-called {\em target space} of the fields $\phi^I$.
Such a system is called a nonlinear $\sigma$ model, harmonic map
or wave map.
The fields $\phi^I$ and metric $G_{IJ}$ are dimensionless,
and this allows the introduction of dimensionless parameters in the
system, which cannot be asymptotically neglected using the arguments of
subsection \ref{section:universalityclasses}.
(Compare with the potential $V(\phi^I)$, which has dimensions
(length)$^{-2}$, and hence requires dimensionful parameters).

The case $N=1$ gives nothing new, and so Hirschmann and Eardly
(HE from now on) studied the $N=2$ case with a target manifold with
constant curvature \cite{HE3}, proportional to a real dimensionless
constant $\kappa$. Using a single complex coordinate $\phi$ 
the action of the system can be written as
\begin{equation}
\int d^4x \sqrt{g} \left[ \frac{R}{2} 
   - \frac{|\nabla\phi|^2}{(1-\kappa|\phi|^2)^2} \right].
\end{equation}
For $\kappa\ge 0$ this system is equivalent to the problem of a
real massless scalar field coupled to Brans-Dicke (BD) gravity (with
the BD coupling constant given by $8\omega_{BD}=-12+ \kappa^{-1}$).
Liebling and Choptuik \cite{LieblingChoptuik} have shown that there is
a smooth transition in the BD system from DSS criticality for
low $\kappa$ (the flat target space case $\kappa=0$ is equivalent to a
self-gravitating complex massless scalar field,
whose critical solution is the original DSS spacetime found by
Choptuik) to CSS criticality for larger $\kappa$ (the case $\kappa=1$
is equivalent to the axion-dilaton system, and has been shown to
display CSS criticality in \cite{HamadeHorneStewart}).

Generalizing their previous results for $\kappa=0$ \cite{HE1,HE2},
HE constructed for each $\kappa$ a CSS solution
based on the ansatz $\phi(\tau,x)=e^{i\omega\tau}\phi(x)$
for the critical scalar field. Studying its perturbations
HE concluded that this solution is critical for
$\kappa>0.0754$, but has three unstable modes for $\kappa<0.0754$ and
even more for $\kappa<-0.28$. Below 0.0754 a DSS solution takes over,
as shown in the simulations of Liebling and Choptuik
\cite{LieblingChoptuik}, and HE conjectured that the transition is
a Hopf bifurcation, such that the DSS cycle smoothly shrinks with
growing $\kappa$, collapsing onto the CSS solution at the transition
and then disappearing with a finite value of the echoing period
$\Delta$.

The close relation between the CSS and DSS critical solutions is also 
manifest in the construction of their global structures. In particular,
the results of \cite{HE2} and \cite{EardleyHirschmannHorne} for the CSS
$\kappa=0$ and $\kappa=1$ solutions respectively show that the Cauchy
horizon of the singularity is almost but not quite flat, exactly as
was the case with the Choptuik DSS spacetime (see subsection
\ref{section:globalstructure}).


\subsubsection{Spherical Einstein-$SU(2)$ sigma model}

This is an $N=3$ sigma model, and it also displays a transition
between CSS and DSS criticality, but this is a totally different
type of transition, in particular showing a divergence in the
echoing period $\Delta$ \cite{Aichelburg1}.

In a reduction to spherical symmetry, the effective action is
\begin{equation}
\int d^4x \sqrt{g} \left[\frac{R}{2}- \eta
\left( |\nabla\phi|^2 + \frac{2\sin^2\phi}{r^2} \right)\right],
\end{equation}
where $r$ is the area radius, and the coupling constant $\eta$ is
dimensionless.
It has been shown (numerically \cite{BizonWassermann} and then
analytically \cite{BizonWassermann2}) that for $0\le\eta<1/2$,
there is an infinite sequence of CSS solutions $\phi_n$ labelled by
a nodal number $n$, and having $n$ growing modes. (The case $\eta=0$,
in which the sigma field decouples from gravity, will be revisited
below.)
The $n$-th solution is always regular in the past light cone of the
singularity, but is regular up to the future light cone only for
$\eta<\eta_n$ where $\eta_0\simeq 0.0688$, $\eta_1\simeq 0.1518$ and
$\eta_n<\eta_{n+1}<1/2$. For larger couplings an apparent horizon
develops and the solution cannot be smoothly continued.
These results suggest that $\phi_0$ is a stable naked singularity
for $\eta<\eta_0$, and $\phi_1$ acts as a critical solution between
naked singularity formation and dispersal for $\eta<\eta_0$ and between
black hole formation and dispersal for $\eta_0<\eta<\eta_1$.
The numerical experiments agree with this scenario in the range
$0\le\eta < 0.14$. Other CSS solutions of this system are investigated
in \cite{BizonSzybkaWassermann}, and the possibility of chaos in
\cite{Szybka}. 

Aichelburg and collaborators \cite{Aichelburg1,LechnerPhD} have shown
that for $\eta\gtrsim 0.2$ there is clear DSS type II criticality
at the black hole threshold. The period $\Delta$ depends on
$\eta$, monotonically decreasing towards an asymptotic value for
$\eta\to\infty$.
Interesting new behavior occurs in the intermediate range
$0.14<\eta<0.2$ that lies between clear CSS and clear DSS.
With decreasing $\eta$ the overall DSS includes episodes of
approximate CSS \cite{Aichelburg2}, of increasing length
(measured in the log-scale time $\tau$). As $\eta\to\eta_c\simeq
0.170$ from above the duration of the CSS epochs, and hence the
overall DSS period $\Delta$ diverges. For $0.14<\eta<0.17$ time
evolutions of initial data near the black hole threshold no longer
show overal DSS, but they still show CSS episodes. Black hole mass
scaling is unclear in this regime.

It has been conjectured that this transition from CSS to DSS can be
interpreted, in the language of the theory of dynamical systems, as
the infinite-dimensional analogue of a three-dimensional Shil'nikov
bifurcation \cite{Lechner}. High-precision numerics in
\cite{Aichelburg4} further supports this picture: for $\eta>\eta_c$ a
codimension-1 CSS solution coexists in phase space with a
codimension-1 DSS attractor such that the (one-dimensional) unstable
manifold of the DSS solution lies on the stable manifold of the CSS
solution.  For $\eta$ close to $\eta_c$ the two solutions are close and
the orbits around the DSS solution become slower because they spend
more time in the neighbourhood of the CSS attractor. A linear
stability analysis predicts a law $\Delta\simeq -\frac{2}{\lambda}
\log(\eta-\eta_c)+b$ for some constant $b$, where $\lambda$ is the
Lyapunov exponent of the CSS solution. For $\eta=\eta_c$ both
solutions touch and the DSS cycle dissapears.


\subsubsection{Einstein-Yang-Mills}

Choptuik, Chmaj and Bizo\'n \cite{ChoptuikChmajBizon} have found
both type I and type II critical collapse in the spherical
Einstein--Yang-Mills system with SU(2) gauge potential, restricting
to the purely magnetic case, in which the matter is described
by a single real scalar field. The situation
is very similar to that of the massive scalar field, and now the
critical solutions are the well-known static $n=1$ Bartnik-Mckinnon
solution \cite{BartnikMcKinnon} for type I and a DSS solution 
(later constructed in \cite{Gundlach_EYM}) for type II.
In both cases the black holes produced in the supercritical regime
are Schwarzschild black holes with zero Yang-Mills field strength,
but the final states (and the dynamics leading to them) can be
distinguished by the value of the Yang-Mills final gauge potential at
infinity, which can take two values, corresponding to two distinct
vacuum states.

Choptuik, Hirschmann and Marsa \cite{ChoptuikHirschmannMarsa} have
investigated the boundary in phase space between formation of those
two types of black holes, using a code that can follow the time
evolutions for long after the black hole has formed. This is a new
``type III'' phase transition whose critical solution is an unstable
static black hole with Yang-Mills hair \cite{Bizon0,VolkovGaltsov},
which collapses to a hairless Schwarzschild black hole with either
vacuum state of the Yang-Mills field, depending on the sign of its one
growing perturbation mode. This ``colored'' black hole is actually a
member of a 1-parameter family parameterized by its apparent
horizon radius and outside the horizon it approaches the corresponding
BM solution. When the horizon radius approaches zero the three critical
solutions meet at a ``triple point''. What happens there deserves
further investigation.

Millward and Hirschmann \cite{Millward} have further coupled a Higgs
field to the Einstein-Yang-Mills system. New possible end states
appear: regular static solutions, and stable hairy
black holes (different from the colored black holes referred to above).
Again there are type I or type II critical phenomena depending on the
initial conditions.

It is known that the spherical critical solutions within the magnetic
ansatz become more unstable when other components of the gauge field
are taken into account, and so they will not be critical in the general
case.


\subsubsection{Vacuum 4+1}

Bizo\'n, Chmaj and Schmidt \cite{BizonChmajSchmidt} have found a way
of constructing asymptotically flat vacuum spacetimes in 4+1
dimensions which are spherically symmetric while containing
gravitational waves (Birkhoff's theorem does not hold in more than 3+1
dimensions).  Recall that in (3+1 dimensional) Bianchi IX cosmology the
manifold is $M^1\times S^3$ where the $S^3$ is equipped with an
$SU(2)$ invariant (homogeneous but anisotropic) metric
\begin{equation}
ds^2 = L_1^2 \sigma_1^2 + L_2^2 \sigma_3^2 + L_3^2 \sigma_3^2,
\end{equation}
where the $\sigma_i$ are the $SU(2)$ left-invariant 1-forms, and the
$L_i$ are functions of time only. Similarly, the spacetimes of
Bizo\'n, Chmaj and Schmidt are of the product form $M^2\times S^3$
where the $L_i$ now depend only on $r$ and $t$. This gives rise to
nontrivial
dynamics, including a threshold between dispersion and black hole
formation. With the additional $U(1)$ symmetry $L_1=L_2$ there is only
one dynamical degree of freedom. At the black hole threshold, type II
critical phenomena are seen with $\Delta\simeq 0.49$ and $\gamma\simeq
0.3289$. In evolutions with the general ansatz where all $L_i$ are
different, the $U(1)$ symmetry is recovered dynamically
\cite{BizonChmajSchmidt2}. Furthermore, the fact that each
$U(1)$-symmetric solution exists in 3 copies gives rise to additional
critical surfaces, and this lead to the prediction of 2-mode-unstable
solutions, which was verified numerically, and the conjecture that
there is a countable family of DSS solutions with $n$ unstable modes
\cite{BizonChmajSchmidt2}. A similar ansatz can be made in other odd
spacetime dimensions, and in 8+1 dimensions type II critical behaviour
is again observed \cite{Bizonetal2005}.


\subsubsection{Scalar field collapse in 2+1}

Spacetime in 2+1 dimensions is flat everywhere where there is no
matter, so that gravity is not acting at a distance in the usual
way. There are no gravitational waves, and black holes can only be
formed in the presence of a negative cosmological constant.
(See \cite{Carlip} for a review.)

Scalar field collapse in circular symmetry was investigated
numerically by Pretorius and Choptuik \cite{PretoriusChoptuik}, and
Husain and Olivier \cite{HusainOlivier}. In a regime where the
cosmological constant is small compared to spacetime curvature they
find type II critical phenomena with a universal CSS critical
solution, and $\gamma=1.20\pm 0.05$ \cite{PretoriusChoptuik}. The
value $\gamma\simeq 0.81$ \cite{HusainOlivier} appears to be less
accurate.

Looking for the critical solution in closed form, Garfinkle
\cite{Garfinkle2+1} found a countable family of exact spherically
symmetric CSS solutions for a massless scalar field with $\Lambda=0$,
but his results remain inconclusive. The $q=4$ solution appears to
match the numerical evolutions inside the past light cone, but its
past light cone is also an apparent horizon. The $q=4$ solution has
three growing modes although the top one would give $\gamma=8/7\simeq
1.14$ if only the other two could be ruled out
\cite{GarfinkleGundlach2+1}. An attempt at this
\cite{HirschmannWangWu} seems unmotivated. At the same time, it is
possible to embed the $\Lambda=0$ solutions into a family of
$\Lambda<0$ ones \cite{ClementFabbri1,ClementFabbri2,Cavaglia}, which
can be constructed along the lines of
Sec.~\ref{section:universalityclasses}, so that Garfinkle's solution
could be the leading term in an expansion in
$e^{-\tau}(-\Lambda)^{-1/2}$.


\subsubsection{Scalar field collapse in higher dimensions}

Critical collapse of a massless scalar field in spherical symmetry in
5+1 spacetime dimensions was investigated in
\cite{GarfinkleCutlerDuncan}. Results are similar to 3+1 dimensions,
with a DSS critical solution and mass scaling with $\gamma\simeq
0.424$. Birukou et al \cite{Birukou, Birukou2} have developed a code
for arbitrary spacetime dimension. They confirm known results in 3+1
($\gamma\simeq 0.36$) and 5+1 ($\gamma\simeq 0.44$) dimensions, and
investigate 4+1 dimensions. Without a cosmological constant they find
mass scaling with $\gamma\simeq 0.41$ for one family of initial data
and $\gamma\simeq 0.52$ for another. They see wiggles in the $\ln M$
versus $\ln(p-p*)$ plot that indicate a DSS critical solution, but
have not investigated the critical solution directly. With a negative
cosmological constant and the second family, they find
$\gamma=0.49$. \cite{BlandKunstatter} has made a more precise
determination: $\gamma=0.4131\pm 0.0001$.
This was motivated by an attempt to explain this exponent using
an holographic duality between the strong coupling regime of 4+1 gravity
and the weak coupling regime of 3+1 QCD
\cite{Alvarez-Gaume}, which had predicted $\gamma=0.409552$.

Kol \cite{Kol} relates a solution that is related to the Choptuik
solution to a variant of the critical solution in the black-string
black hole transition, and claims to obtain analytic estimates for
$\gamma$ and $\Delta$. This has motivated a numerical determination of
$\gamma$ and $\Delta$ for the spherical massless scalar field in
noninteger dimension up to 14 \cite{SorkinOren,BlandPreston}.


\subsubsection{Other systems obeying wave equations}

Choptuik, Hirschmann and Liebling \cite{ChopHirschLieb} have presented
perturbative indications that the static solutions found by van Putten
\cite{vanPutten} in the vacuum Brans-Dicke system are critical
solutions. They have also performed full numerical simulations, but
only starting from small deviations with respect to those solutions.

Ventrella and Choptuik \cite{VentrellaChoptuik} have performed numerical
simulations of collapse of a massless Dirac field in a special state:
an incoherent sum of two independent left-handed zero-spin fields
having opposite orbital angular momentum. This is prepared so that the
total distribution of energy-momentum is spherically symmetric. The
freedom in the system is then contained in a single complex scalar field
obeying a modified linear wave equation in spherical symmetry. There
are clear signs of CSS criticality in the metric variables, and the
critical complex field exhibits a phase of the form $e^{i\omega\tau}$
for a definite $\omega$ (the Hirschmann and Eardly ansatz for the
complex scalar field critical solutions), which can be considered as a
trivial form of DSS.

Garfinkle, Mann and Vuille \cite{GarfinkleMannVuille} have found
coexistence of types I and II criticality in the spherical collapse
of a massive vector field (the Proca system), the scenario being
almost identical to that of a massive scalar field. In the self-similar
phase the collapse amplifies the longitudinal mode of the Proca field
with respect to its transverse modes, which become negligible, and
the critical solution is simply the gradient of the Choptuik
DSS spacetime.

Sarbach and Lehner \cite{SarbachLehner} find type I critical behaviour
in $q+3$-dimensional spacetimes with $U(1)\times SO(q+1)$ symmetry in
Einstein-Maxwell theory at the threshold between dispersion and
formation of a black string.



\begin{table}

\caption{
\label{table:mattermodels}
Critical collapse in spherical symmetry}

\bigskip

\begin{tabular}{| l | c | c | c | c | c |}

\hline\hline

Matter & Type & Collapse  & Critical & Perturbations \\ 
& & simulations & solution & of crit. soln.\\

\hline\hline

Perfect fluid $p=k\rho$ & II & \cite{EvansColeman,NeilsenChoptuik} &
 CSS \cite{EvansColeman,Maison,NeilsenChoptuik} 
& \cite{Maison,KoikeHaraAdachi2,Gundlach_nonspherical,Gundlach_critfluid2} \\

Vlasov & I? & \cite{ReinRendallSchaeffer,OlabarrietaChoptuik} &
\cite{vlasov1} & \\
\hline

Real scalar field: &&&& \\
 
-- massless, min. coupled & II & \cite{Choptuik91,Choptuik92,Choptuik94} &

DSS \cite{Gundlach_Chop1} 
& \cite{Gundlach_Chop2,MartinGundlach} \\

-- massive & I & \cite{BradyChambersGoncalves} 
& oscillating \cite{SeidelSuen} & \\

& II & \cite{Choptuik94}
& DSS \cite{HaraKoikeAdachi,GundlachMartin} 
& \cite{HaraKoikeAdachi,GundlachMartin} \\ 

-- conformally coupled & II &
\cite{Choptuik92} & DSS  &   \\

-- 4+1 & II & \cite{Birukou} & & \\

-- 5+1 & II & \cite{GarfinkleCutlerDuncan} & & \\

\hline

Massive complex scalar field & I, II & \cite{HawleyChoptuik} &
\cite{SeidelSuen}& \cite{HawleyChoptuik} \\

Massless scalar electrodynamics & II & \cite{HodPiran_charge} &
DSS \cite{GundlachMartin} &
\cite{GundlachMartin} \\ 

\hline

Massive vector field & II & \cite{GarfinkleMannVuille} &
 DSS \cite{GarfinkleMannVuille} & \cite{GarfinkleMannVuille} \\

Massless Dirac & II & \cite{VentrellaChoptuik} & CSS \cite{VentrellaChoptuik} &  \\

Vacuum Brans-Dicke & I & \cite{ChopHirschLieb} & static 
\cite{vanPutten} & 
\cite{ChopHirschLieb}\\

\hline

2-d sigma model &&&& \\

-- complex scalar ($\kappa=0$) & II & \cite{Choptuik_pc} &
DSS \cite{Gundlach_Chop2}  &
\cite{Gundlach_Chop2} \\

-- axion-dilaton ($\kappa=1$) & II & 
\cite{HamadeHorneStewart} & 
CSS \cite{EardleyHirschmannHorne,HamadeHorneStewart} & 
\cite{HamadeHorneStewart} \\

-- scalar-Brans-Dicke ($\kappa>0$) & II & \cite{LieblingChoptuik,Liebling}
& CSS, DSS & \\

-- general $\kappa$ including $\kappa<0$ & II & & CSS, DSS \cite{HE3} & \cite{HE3} \\

\hline
 
$SU(2)$ Yang-Mills & I & \cite{ChoptuikChmajBizon} &
static \cite{BartnikMcKinnon} & \cite{LavrelashviliMaison} \\

& II & \cite{ChoptuikChmajBizon} & DSS \cite{Gundlach_EYM} 
& \cite{Gundlach_EYM} \\

& ``III'' & \cite{ChoptuikHirschmannMarsa} &
colored BH \cite{Bizon0,VolkovGaltsov} &
\cite{StraumannZhou,VolkovBrodbeckLavrelashviliStraumann,BizonChmaj3} \\

$SU(2)$ Yang-Mills-Higgs & (idem) & \cite{Millward} &
(idem)  & \\

$SU(2)$ Skyrme model & I & \cite{BizonChmaj} & static \cite{BizonChmaj}
& \cite{BizonChmaj} \\

& II & \cite{BizonChmajTabor}  

& DSS \cite{BizonChmajTabor}  & \\

$SO(3)$ Mexican hat & II & \cite{Liebling2} & DSS & \\

\hline

\hline

\end{tabular}
\end{table}



\subsection{Perfect fluid matter}
\label{section:perfectfluid}

\subsubsection{Spherical symmetry}

Evans and Coleman \cite{EvansColeman} performed the first simulations
of critical collapse with a perfect fluid with equation of state
$p=k\rho $ (where $\rho$ is the energy density and $p$ the pressure)
for $k=1/3$ (radiation), and found a CSS critical solution with a
mass-scaling critical exponent $\gamma\approx 0.36$.  Koike, Hara and
Adachi \cite{KoikeHaraAdachi} constructed that critical solution and
its linear perturbations from a CSS ansatz as an eigenvalue problem,
computing the critical exponent to high precision.  Independently,
Maison \cite{Maison} constructed the regular CSS solutions and their
linear perturbations for a large number of values of $k$, showing for
the first time that the critical exponents were model-dependent. As
Ori and Piran before \cite{OriPiran,OriPiran2}, he claimed that there
are no regular CSS solutions for $k>0.89$, but Neilsen and Choptuik
\cite{NeilsenChoptuik,NeilsenChoptuik2} have found CSS critical
solutions for all values of $k$ right up to 1, both in collapse
simulations and by making a CSS ansatz. The difficulty comes from a
change in character of the sonic point, which becomes a nodal point
for $k>0.89$, rather than a focal point, making the ODE problem
associated with the CSS ansatz much more difficult to solve.
Harada \cite{Harada2} has also found that the critical solution
becomes unstable to a ``kink'' (discontinuous at the sonic point of
the background solution) mode for $k>0.89$, but because it is not
smooth it does not seem to have any influence on the
numerical simulations of collapse. On the other hand, the limit $k\to 1$
leading to the stiff equation of state $p=\rho$ is singular in that
during evolution the fluid 4-velocity can become spacelike and the
density $\rho$ negative. The stiff fluid equations of motion are in fact
equivalent to the massless-scalar field, but the critical solutions
can differ, dependending on how one deals with the issue of negative
density \cite{BradyChoptuikGundlachNeilsen}. Summarizing, it is
possible to construct the Evans-Coleman CSS critical (codimension 1)
solution for all values $0<k<1$. This solution can be identified in
the general classification of CSS perfect-fluid solutions as the
unique spacetime that is analytic at the center and at the sound cone,
is ingoing near the center, and outgoing everywhere else
\cite{CarrColey2,CarrColeyGoliathNilssonUggla,CarrGundlach}.
There is even a Newtonian counterpart of the critical solution:
the Hunter(a) solution \cite{HaradaMaeda}.

$p=k\rho$ is the only equation of state compatible with exact CSS
(homothetic) solutions for perfect fluid collapse \cite{CahillTaub}
and therefore we might think that other equations of state would not
display critical phenomena, at least of type II. Neilsen and Choptuik
\cite{NeilsenChoptuik} have given evidence that for the ideal gas
equation of state (from now on EOS)
$p=k\rho_0\epsilon$ (where $\rho_0$ is the rest mass density and
$\epsilon$ is the internal energy per rest mass unit) the black
hole threshold also contains a CSS attractor, and that it coincides
with the CSS exact critical solution of the ultrarelativistic case
with the same $k$. This is interpreted a posteriori as a sign that the
critical CSS solution is highly ultrarelativistic
$\rho=(1+\epsilon)\rho_0 \simeq \epsilon\rho_0\gg \rho_0$, and hence
rest mass is irrelevant. Novak \cite{Novak} has also shown in the case
$k=1$, or even with a more general tabulated EOS, that type II
critical phenomena can be found by velocity-induced perturbations of
static TOV solutions. A thorough and much more precise analysis by
Noble and Choptuik \cite{Noble, NobleChoptuik} of the possible collapse
scenarios of
the stiff $k=1$ ideal gas has confirmed this surprising result, and
again the critical solution (and hence the critical exponent) is that
of the ultrarelativistic limit problem.  Parametrizing, as usual, the
TOV solutions by the central density $\rho_c$, they find that for
low-density initial stars it is not possible to form a black hole by
velocity-induced collapse; for intermediate initial values of
$\rho_c$, it is possible to induce type II criticality for large
enough velocity perturbations; for large initial central densities
they always get type I criticality, as we might have anticipated.

Noble and Choptuik \cite{Noble} have also investigated the evolution
of a perfect fluid interacting with a massless scalar field indirectly
through gravity. By tuning of the amplitude of the pulse it is
possible to drive a fluid star to collapse.  For massive stars type I
criticality is found, in which the critical solution oscillates around
a member of the unstable TOV branch. For less massive stars a large
scalar amplitude is required to induce collapse, and the black hole
threshold is always dominated by the scalar field DSS critical
solution, with the fluid evolving passively.

\subsubsection{Nonspherical perturbations}

Non-spherically symmetric perturbations around the spherical critical
solution for the perfect fluid can be used to study angular momentum
perturbatively. All nonspherical perturbations of the perfect fluid
critical solution decay for equations of state $p=k\rho$ with $k$ in
the range $1/9<k<0.49$ \cite{Gundlach_nonspherical}, and so the
spherically symmetric critical solution is stable under small
deviations from spherical symmetry. Infinitesimal angular momentum is
carried by the axial parity perturbations with angular dependence
$l=1$. From these two facts one can derive the angular momentum
scaling law at the black hole threshold
\cite{Gundlach_angmom,Gundlach_critfluid2}
\begin{equation}
\label{CSS_L}
L \sim (p-p_*)^\mu,
\end{equation}
which should be valid in the range $1/9<k<0.49$. The angular momentum
exponent $\mu(k)$ is related to the mass exponent $\gamma(k)$ by
\begin{equation}
\label{muofk}
\mu(k) = \Big(2+\lambda_1(k)\Big)\gamma(k) .
\end{equation}
where 
\begin{equation}
\label{lambda1}
\lambda_1(k)={9k-1\over 3k+3}
\end{equation}
is the growth or decay rate of the dominant $l=1$ axial
perturbation mode. In particular for the value $k=1/3$, where
$\gamma\simeq 0.3558$, $\mu=(5/2)\gamma\simeq 0.8895$. 

Ori and Piran \cite{OriPiran2} have pointed out that there exists a
CSS perfect fluid solution for $0<k<0.036$ generalizing the
Larson-Penston solution of Newtonian fluid collapse, and which
has a naked singularity for $0<k<0.0105$. Harada and Maeda
\cite{Harada,HaradaMaeda} have shown that this solution has no growing
perturbative modes in spherical symmetry and hence a naked singularity
becomes a global attractor of the evolution for the latter range of
$k$. This is also true in the limit $k=0$, which can be considered as
the Newtonian limit \cite{HaradaMaeda2, HaradaMaeda3}. 
Their result has been confirmed with very high precision
numerics by Snajdr \cite{Snajdr}.
This seems to violate cosmic censorship, as generic spherical initial
data would create a naked singularity. However, the exact result
(\ref{lambda1}) holds for any regular CSS spherical perfect fluid
solution, and so all such 
solutions with $k<1/9$ have at least one unstable nonspherical
perturbation. Therefore the naked singularity is unstable
to infinitesimal perturbations with angular momentum when one lifts the
restriction to spherical symmetry.

In the early universe, quantum fluctuations of the metric and
matter can be important, for example providing the seeds of galaxy
formation. Large enough fluctuations will collapse to form primordial
black holes. As large quantum fluctuations are exponentially more
unlikely than small ones, $P(\delta)\sim e^{-\delta^2}$, where
$\delta$ is the density contrast of the fluctuation, one would expect
the spectrum of primordial black holes to be sharply peaked at the
minimal $\delta$ that leads to black hole formation, giving rise to
critical phenomena \cite{NiemeyerJedamzik}. See also
\cite{GreenLiddle,Yokoyama}. 

An approximation to primordial black hole formation is a spherically
symmetric distribution of a radiation gas ($p=\rho/3$) with
cosmological rather than asymptotically flat boundary conditions. In
\cite{NiemeyerJedamzik,NiemeyerJedamzik2} type II critical phenomena
were found, which would imply that the mass of primordial black holes
formed are much smaller than the naively expected value of the mass
contained within the Hubble horizon at the time of collapse. The
boundary conditions and initial data were refined in
\cite{HawkeStewart,MuscoMiller}, and a minimum black hole mass of
$\sim 10^{-4}$ of the horizon mass was found, due to matter accreting
onto the black hole after strong shock formation.


\subsection{Collisionless matter}

A cloud of collisionless particles can be described by the Vlasov
equation, i.e., the Boltzmann equation without collision term.
This matter model differs from field theories by having a much larger
number of matter degrees of freedom: The matter content is described
by a statistical distribution $f(x^\mu, p_\nu)$ on the point particle
phase space, instead of a finite number of fields $\phi(x^\mu)$.
When restricted to spherical symmetry, individual particles move
tangentially as well as radially, and so individually have angular
momentum, but the stress-energy tensor averages out to a spherically
symmetric one, with zero total angular momentum. The distribution
$f$ is then a function $f(r,t,p^r,L^2)$ of radius, time, radial
momentum and (conserved) angular momentum.

Several numerical simulations of critical collapse of collisionless
matter in spherical symmetry have been published to date, and
remarkably no type II scaling phenomena has been discovered.
Indications of type I scaling have been found, but these do not
quite fit the standard picture of critical collapse. Rein et al.\
\cite{ReinRendallSchaeffer} find that black hole formation turns on
with a mass gap that is a large part of the ADM mass of the initial
data, and this gap depends on the initial matter condition.  No
critical behavior of either type I or type II was observed.
Olabarrieta and Choptuik \cite{OlabarrietaChoptuik} find evidence of a
metastable static solution at the black hole threshold, with type I
scaling of its life time as in Eq.\ (\ref{typeIscaling}).  However,
the critical exponent depends weakly on the family of initial data,
ranging from 5.0 to 5.9, with a quoted uncertainty of 0.2.
Furthermore, the matter distribution does not appear to be universal,
while the metric seems to be universal up to an overall rescaling,
so that there appears to be no universal critical solution. More
precise computations by Stevenson and Choptuik \cite{Stevenson}, using
finite volume HRSC methods, have confirmed the existence of static
intermediate solutions and non-universal scaling with exponents
ranging now from 5.27 to 11.65.

Mart\'\i n-Garc\'\i a and Gundlach \cite{vlasov1} have constructed
a family of CSS spherically symmetric solutions for massless particles
that is generic by function counting. There are infinitely many
solutions with different matter configurations but the same
stress-energy tensor and spacetime metric, due to the existence of an
exact symmetry: two massless particles with energy-momentum $p^\mu$
in the solution can be replaced by one particle with $2p^\mu$.
A similar result holds for the perturbations. As the growth exponent
$\lambda$ of a perturbation mode can be determined from the metric
alone, this means that there are infinitely many perturbation modes
with the same $\lambda$. If there is one growing perturbative mode,
there are infinitely many. Therefore a candidate critical solution
(either static or CSS) cannot be isolated or have only one growing mode.
This argument rules out the existence of both type I and
type II critical phenomena (in their standard form, i.e., including
universality) for massless particles in the complete system, but
some partial form of criticality could still be found by restricting
to sections of phase space in which that symmetry is broken, for
example by prescribing a fixed form for the dependence of the
distribution function $f$ on angular momentum $L$, as those
numerical simulations have done.

A recent investigation of Andr\'easson and Rein \cite{AndreassonRein}
with massive particles has confirmed again the existence of a mass gap
and the existence of metastable static solutions at the black hole
threshold, though there is no estimation of the scaling of their
life-times. More interestingly, they show that the subcritical regime
can lead to either dispersion or an oscillating steady state depending
on the binding energy of the system. They also conclude, based on
perturbative arguments, that there cannot be an isolated universal
critical solution.

More numerical work is still required, but 
current evidence suggests that there are no type II critical
phenomena, and that there is a continuum of critical solutions in type
I critical phenomena and hence only limited universality.


\subsection{Criticality in singularity formation without 
gravitational collapse}

It is well known that the Yang-Mills field does not form singularities
from smooth initial conditions in 3+1 dimensions
\cite{EardleyMoncrief}, but Bizo\'n and Tabor \cite{BizonTabor} have
shown singularity formation in 4+1 (the critical dimension for this
system from the point of view of energy scaling arguments) and 5+1
(the first supercritical dimension). In 5+1 there is countable family
$W_n$ of CSS solutions with $n$ unstable modes, such that $W_1$ acts
as a critical solution separating singularity ($W_0$) formation from
dispersal to infinity.  In 4+1 there are no self-similar solutions and
the formation of singularities seems to proceed through adiabatic
shrinking of a static solution.

Completely parallel results can be found for wave maps, for which
the critical dimension is 2+1.
For the wave map from 3+1 Minkowski to the 3-sphere
Bizo\'n \cite{Bizon1} has shown that there is a countable family of
regular (before the Cauchy horizon) CSS solutions labeled by a nodal
number $n\ge 0$, such that each solution has $n$ unstable modes.
Simulations of collapse in spherical symmetry
\cite{BizonChmajTabor2,LieblingHirschmannIsenberg}
and in 3D \cite{Liebling3D} show that $n=0$ is a global
attractor and the $n=1$ solution is the critical solution. See also
\cite{DonningerAichelburg1,DonningerAichelburg2} for 
computations of the largest perturbation-eigenvalues of $W_0$ and $W_1$.
Again, for the wave map from
2+1 Minkowski to the 2-sphere generic singularity formation proceeds
through adiabatic shrinking of a static solution
\cite{BizonChmajTabor3}.

These results have led to the suggestion in \cite{BizonTabor} that
criticality (in the sense of the existence of a codimension-1 solution
separating evolution towards qualitatively different end states) could
be a generic and robust feature of evolutionary PDE systems in
supercritical dimensions, and not an effect particular of gravity.

Garfinkle and Isenberg \cite{GarfinkleIsenberg} examine the threshold
between the round endstate and pinching off in Ricci flow for a familiy
of spherically symmetric geometries on $S^3$. They have found
intermediate approach to a special ``javelin'' geometry, but have
not investigated whether this is universal.

A scaling of the shape of the event horizon at the
moment of merger in binary black hole mergers is noted in
\cite{Caveny}, but is really a kinematic effect.


\subsection{Analytic studies and toy models}


\subsubsection{Exact solutions of Einstein--Klein-Gordon}

A number of authors have explored the possibility of finding critical
phenomena with CSS (rather than DSS) massless scalar critical solutions.
The Roberts 1-parameter family of 3+1 solutions \cite{Roberts}
has been analyzed along this line in
\cite{Oshiro_Roberts, Brady_Roberts, WangOliveira, Kiem}. This family
contains black holes whose masses (with a suitable matching to an
asymptotically flat solution) scale as $(p-1)^{1/2}$ for $p\gtrsim 1$,
but such a special family of solutions
has no direct relevance for collapse from generic data.
Its generalization to other dimensions has been considered in
\cite{Frolov2}.
A fully analytic construction of all (spherical and nonspherical) linear
perturbations of the Roberts solution by Frolov \cite{Frolov, Frolov3}
has shown that there is a continuum of unstable spherical modes
filling a sector of the complex plane with $Re\lambda\ge 1$, so that
it cannot be a critical solution. Interestingly, all nonspherical
perturbations decay. 

Frolov \cite{Frolov4} has also suggested that the critical ($p=1$)
Roberts solution,
which has an outgoing null singularity, plus its most rapidly growing
(spherical) perturbation mode would evolve into the Choptuik solution,
which would inherit the oscillation in $\tau$ with a period $4.44$ of
that mode.

A similar transition within a single 1-parameter family of
solutions has been pointed out in \cite{Oliveira} for the Wyman
solution \cite{Wyman}.

Hayward \cite{Hayward,Haywarderratum} and Clement and Fabbri
\cite{ClementFabbri1,ClementFabbri2} have also proposed critical
solutions with a null singularity, and have attempted to construct
black hole solutions from their linear perturbations. This is probably
irrelevant to critical collapse, as the critical spacetime does not
have an outgoing null singularity. Rather, the singularity is naked
but first appears in a point. The future light cone of that point is
not a null singularity but a Cauchy horizon with finite
curvature.

Other authors have attempted analytic approximations to the Choptuik
solution.  Pullin \cite{Pullin_Chop} has suggested describing critical
collapse approximately as a perturbation of the Schwarzschild
spacetime. Price and Pullin \cite{PricePullin} have approximated the
Choptuik solution by two flat space solutions of the scalar wave
equation that are matched at a ``transition edge'' at constant
self-similarity coordinate $x$. The nonlinearity of the gravitational
field comes in through the matching procedure, and its details are
claimed to provide an estimate of the echoing period $\Delta$.


\subsubsection{Toy models}

Birmingham and Sen \cite{BirminghamSen} considered the formation of a
black hole from the collision of two point particles of equal
mass in 2+1 gravity. Peleg and Steif \cite{PelegSteif} have
investigated the collapse of a dust ring. In both cases the mass
of the black holes is a known function of the parameters of the
initial condition, giving a ``critical exponent'' 1/2, but no
underlying self-similar solution is involved.

Frolov \cite{FrolovV} and Frolov, Larsen and Christensen
\cite{FrolovVetal} consider a stationary 2+1-dimensional Nambu-Goto
membrane held fixed at infinity in a stationary 3+1-dimensional black
hole background spacetime. The induced 2+1 metric on the membrane can
have wormhole, black hole, or Minkowski topology. The critical
solution between Minkowski and black hole topology has 2+1 CSS. The
mass of the apparent horizon of induced black hole metrics scales with
$\gamma=2/3$, superimposed with a wiggle of period $3\pi/\sqrt{7}$ in
$\ln p$. The mass scaling is universal with respect to different
background black hole metrics, as they can be approximated by Rindler
space in the mass scaling limit.

Horowitz and Hubeny \cite{HorowitzHubeny} and Birmingham
\cite{BirminghamAdS} have attempted to calculate the critical exponent
in toy models from the adS-CFT correspondence.

Burko \cite{Burko3} considers the transition between existence and
non-existence of a null branch of the singularity inside a spherically
symmetric charged black hole with massless scalar field matter thrown
in.

Wang \cite{WangCylindrical} has constructed homothetic cylindrically
symmetric solutions of 3+1 Einstein-Klein-Gordon and studied their
cylindrically symmetric perturbations. It is not clear how these
are related to a critical surface in phase space.


\subsection{Quantum effects}

Type II critical phenomena provide a relatively natural way of
producing arbitrarily high curvatures, where quantum gravity effects
should become important, from generic initial data. Approaching the
Planck scale from above, one would expect to be able to write down a
critical solution that is the classical critical solution
asymptotically at large scales, as an expansion in inverse powers of
the Planck length, see Sec.\ref{section:universalityclasses}.

Black hole evolution in semiclassical gravity has been investigated in
1+1 dimensional models which serve as toy models for spherical
symmetry (see \cite{Giddings_BH} for a review). The black hole
threshold in such models has been investigated in \cite{ChibaSiino},
\cite{AyalPiran}, \cite{StromingerThorlacius}, \cite{Kiem},
\cite{ZhouKirstenYang}, \cite{PelegBoseParker,BoseParkerPeleg}. In some
of these models, the critical exponent is $1/2$ for kinematical reasons.

In \cite{BradyOttewill} a 3+1-dimensional but perturbative approach is
taken. The quantum effects then give an additional unstable mode with
$\lambda=2$. If this is larger than the positive Lyapunov exponent
$\lambda_0$, it will become the dominant perturbation for sufficiently
good fine-tuning, and therefore sufficiently good fine-tuning will
reveal a mass gap.


\section{Beyond spherical symmetry}
\label{section:nonspherical}

Numerical studies of critical collapse should go beyond spherical
symmetry (and in the first instance to axisymmetry) for three reasons:

\begin{itemize}

\item Weak gravitational waves in vacuum general relativity can focus
and collapse. The black hole threshold in this process shows what in
critical phenomena in gravitational collapse is intrinsic to gravity
rather than the matter model. 

\item Black holes are characterised by charge and angular momentum as
well as mass. Angular momentum is the more interesting of the two
because it is again independent of matter, but cannot be studied in
spherical symmetry.

\item Angular momentum resists collapse, but angular momentum in the
initial data is needed to make a black hole with angular
momentum. Therefore it is an interesting question to ask what
happens to the dimensionless ratio $J/M^2$ at the black hole
threshold. 

\end{itemize}

In the following we review what has been done so far. 


\subsection{Perturbative approach to angular momentum}

We have already mentioned that when angular momentum is small, a
critical exponent for $|J|$ can be derived in perturbation
theory. This has been done for the perfect fluid (see
Sec.~\ref{section:perfectfluid}) in first-order perturbation theory
and for the massless scalar field (see
Sec.~\ref{section:scalarnonspherical}), where second-order
perturbation theory in the scalar field is necessary to obtain an
angular momentum perturbation in the stress-energy tensor
\cite{GarfinkleGundlachMartin}. However, neither of the predicted
angular momentum scaling laws has been verified in numerical
evolutions.

For a perfect fluid with equation of state $p=k\rho$ with $0<k<1/9$,
precisely one mode that carries angular momentum is unstable, and
this mode and the known spherical mode are the only two unstable modes
of the spherical critical solution. (Note by comparison that from
dimensional analysis one would not expect an uncharged critical
solution to have a growing perturbation mode carrying charge.)  The
presence of two growing modes of the critical solution is
expected to give rise to interesting phenomena
\cite{Gundlach_scalingfunctions}. Near the critical solution, the two
growing modes compete. $J$ and $M$ of the final black hole are
expected to depend on the distance to the black hole threshold and the
angular momentum of the initial data through universal functions of
{\em one} variable that are similar to ``universal scaling functions''
in statistical mechanics (see also the end of
Sec.~\ref{section:analogy}). While they have not yet been computed,
these functions can in principle be determined from time evolutions of
a single 2-parameter family of initial data, and then determine $J$
and $M$ for all initial data near the black hole threshold and with
small angular momentum. They would extend the simple power-law
scalings of $J$ and $M$ into a region of initial data space with
larger angular momentum.


\subsection{Axisymmetric vacuum gravity}
\label{section:brillwaves}

Abrahams and Evans \cite{AbrahamsEvans} have numerically investigated
black hole formation in axisymmetric vacuum gravity. They write
the metric as
\begin{equation}
\label{axisymmetric}
ds^2=-\alpha^2\,dt^2 + \phi^4\left[e^{2\eta/3}(dr+\beta^r\,dt)^2 +
r^2e^{2\eta/3}(d\theta+\beta^\theta\,dt)^2 +
e^{-4\eta/3}r^2\sin^2\theta\,d\varphi^2\right],
\end{equation}
where the lapse $\alpha$, shift components $\beta^r$ and
$\beta^\theta$, and 3-metric coefficients $\phi$ and $\eta$ are
functions of $r$, $t$ and $\theta$. Axisymmetry limits gravitational
waves to one polarisation out of two, so that there are as many
physical degrees of freedom as in a single wave equation. On the
initial slice, $\eta$ and $K^r_\theta$ are given as free data, and
$\phi$, $K^r_r$, and $K^\varphi_\varphi$ are determined by solving the
Hamiltonian constraint and the two independent components of the
momentum constraint. Afterwards, $\eta$, $K^r_\theta$, $K^r_r$, and
$K^\varphi_\varphi$ are evolved, and only $\phi$ is obtained by
solving the Hamiltonian constraint.  $\alpha$ is obtained by solving
the maximal slicing condition ${K_i}^i=0$ for $\alpha$, and $\beta^r$
and $\beta^\theta$ are obtained from the time derivatives
quasi-isotropic spatial gauge conditions $g_{\theta\theta}=r^2 g_{rr}$
and $g_{r\theta}=0$.

In order to keep their numerical grid as small as possible, Abrahams
and Evans chose their initial data to be mostly ingoing. The two free
functions $\eta$ and $K^r_\theta$ in the initial data were chosen to
have the same functional form they would have in a linearised
gravitational wave with pure $l=2,m=0$ angular dependence. This ansatz
reduced the freedom in the initial data to one free function of
advanced time. A specific peaked function was chosen, and only the
overall amplitude was varied.

Limited numerical resolution allowed Abrahams and Evans to find black
holes with masses only down to $0.2$ of the ADM mass. Even this far
from criticality, they found power-law scaling of the black hole mass,
with a critical exponent $\gamma\simeq 0.36$. The black hole mass was
determined from the apparent horizon surface area, and the frequencies
of the lowest quasi-normal modes of the black hole.  There was
tentative evidence for scale echoing in the time evolution, with
$\Delta\simeq 0.6$, with about three echos seen. Here $\eta$ has the
echoing property $\eta(e^\Delta r,e^\Delta t)=\eta(r,t)$, and the same
echoing property is expected to hold also for $\alpha$, $\phi$,
$\beta^r$ and $r^{-1}\beta^\theta$. In a subsequent paper
\cite{AbrahamsEvans2}, some evidence for universality of the critical
solution, echoing period and critical exponent was given in the
evolution of a second family of initial data, one in which $\eta=0$ at
the initial time. In this family, black hole masses down to $0.06$ of
the ADM mass were achieved.

It is striking that at 14 years later, these results have not yet been
independently verified. An attempt with a 3D numerical relativity code
(but axisymmetric initial data), using free evolution in the BSSN
formulation with maximal slicing and zero shift, to repeat the results
of Abrahams and Evans was not successful \cite{Alcubierre_Brill}. The
reason could be a combination of rather lower resolution than that of
Abrahams and Evans, growing constraint violations, and an
inappropriate choice of gauge. (It is now known that BSSN with maximal
slicing and zero shift is an ill-posed system \cite{bssn3}.)


\subsection{Scalar field}

In 2003, Choptuik, Hirschmann, Liebling and Pretorius
reported on numerical evolutions
at the black hole threshold of an axisymmetric massless scalar field
\cite{Choptuik_axisymmetry}. In axisymmetry with scalar field matter
there is no angular momentum and only one
polarisation of gravitational waves. The slicing condition is maximal
slicing and the spatial gauge, in cylindrical coordinates, is
$g_{zz}=g_{\rho\rho}$, $g_{z\rho}=0$, similar to the gauge used by
Abrahams and Evans. The Hamiltonian constraint is solved at every time
step, and the time derivatives of the spatial gauge conditions are
substituted into the momentum constraints to obtain second-order
elliptic equations for the two shift components. Thus the evolution is
partially constrained. Adaptive mesh-refinement was used in the
numerical time evolution. 

The initial data were either time-symmetric or approximately ingoing, with
the scalar field either symmetric or antisymmetric in $z$. In the
symmetric case, even strongly non-spherical data were attracted to the
known spherical critical solution for the massless scalar
field. Scaling with the known $\gamma$ was observed in the Ricci
scalar. However, with sufficiently good fine-tuning to the black hole
threshold, the approximately spherical region that approaches the
critical solution suffers an $l=2$ (and by ansatz $m=0$) instability
and splits into two new spherical regions which again
approach the critical solution. The spatial separation of the two new
centres is related to the smallest length scale that developed prior
to the branching. There is evidence that with increasing fine-tuning
each of these centres splits again. The antisymmetric initial data
cannot approach a single spherical critical solution but the solution
splits initially into two approximately spherical regions where the
critical solution is approached (up to an overall sign in the scalar
field). The separation of these initial two centres is determined by
the initial data, but there is evidence that they in turn split.

All this is consistent with the assumption that the spherical critical
solution has, besides the known one spherical unstable mode, precisely
one further $l=2$ unstable mode. (Without the restriction to
axisymmetry, if such a mode exists, it would be 5-fold degenerate with
$m=-2,\dots,2$.)  This contradicts the calculation of the perturbation
spectrum in \cite{MartinGundlach}. Choptuik and co-workers do not
state with certainty that the mode they see in numerical evolutions is
a continuum mode, although they have no indication that it is a
numerical artifact. The growth rate of the putative mode is measured
to be $\lambda\simeq 0.1-0.4$, which should be compared with the
growth rate $\lambda\simeq 2.7$ of the spherical mode and the
relatively small decay rate of $\lambda\simeq -0.02$ claimed in
\cite{MartinGundlach} for the least damped mode, which is also an
$l=2$ mode. We observe that the range of $\tau$ in Figs.~6 and 7 of
\cite{Choptuik_axisymmetry} is about 10, and over this range the plot
of the amplitude of the $l=2$ perturbation against the log-scale
coordinate $\tau$ seems equally consistent with linear growth in
$\tau$ as with exponential growth. (In the notation of
\cite{Choptuik_axisymmetry}, $\tau$ denotes proper time and $\tau_*$
the accumulation point of echos, so that the log-scale coordinate
$\tau$ used in this review corresponds to $-\ln(\tau-\tau_*)$ in the
notation of \cite{Choptuik_axisymmetry}.)

An interesting extension was made in
\cite{ChoptuikHirschmannLieblingPretorius} by considering a complex
scalar field giving rise to an axisymmetric spacetime with angular
momentum. The stress-energy tensor of a complex scalar field $\Psi$ is
\begin{equation}
T_{ab}={1\over2}\left(\Psi_{,a}\bar\Psi_{,b}+\bar\Psi_{,a}\Psi_{,b}
- g_{ab}\bar\Psi_{,c}\Psi^{,c}\right)
\end{equation}
An axisymmetric spacetime with azimuthal Killing vector $\xi$ admits a
conserved vector field $T_{ab}\xi^b$, so that the total angular momentum
\begin{equation}
J=\int_\Sigma T_{ab}\xi^a n^b \,dV
\end{equation}
is independent of the Cauchy surface $\Sigma$. With the ansatz
\begin{equation} \label{m1ansatz}
\Psi(\rho,z,t,\varphi)=\Phi(\rho,z,t)e^{im\varphi},
\end{equation}
in adapted coordinates where $\xi=\partial/\partial \varphi$ and with
$m$ an integer, the stress-energy tensor becomes axisymmetric and
hence compatible with the Einstein equations for an axisymmetric spacetime. 

For any $\Sigma$ tangent to $\xi$, in particular a hypersurface
$t={\rm const.}$, the angular momentum density measured by a normal
observer becomes
\begin{equation}
j=T_{ab}\xi^a n^b = -{im\over 2}\left(\Pi\bar\Phi-\bar\Pi\Phi\right)
=m A^2 n^a \delta_{,a},
\end{equation}
where $\Phi=Ae^{i\delta}$ with $\delta$ and $A$ real and
$\Pi=n^a\Phi_{,a}$. By comparison the energy density measured by a normal
observer is
\begin{equation}
\rho=T_{ab}n^an^b=\Pi\bar\Pi+D_a\Phi D^a\bar\Phi,
\end{equation}
where $D_a$ is the derivative operator projected into $\Sigma$. This
means that the ratio of energy density to angular momentum density can
be adjusted arbitrarily in the initial data, including zero angular
momentum for a $\Phi$ that is real (up to a constant phase). On the
other hand, even in the absence of angular momentum a purely real
$\Phi$ obeys a wave equation with an explicit $m^2/\rho^2$ centrifugal
term.  For the same reason, regular solutions must have
$\Phi\sim\rho^m$ on the axis, and there are no spherically symmetric
solutions. Intuitively
speaking, the centrifugal force resisting collapse appears 
unrelated to the angular momentum component of the stress-energy
tensor in a way that differs from what one would expect in rotating
fluid collapse or rotating (non-axisymmetric) vacuum collapse.

For all initial data in numerical evolutions, a critical solution is
approached that is discretely self-similar with log-scale period
$\Delta\simeq 0.42$. (By ansatz this solution is axisymmetric but
spherical symmetry is ruled out and so the critical solution cannot be
the Choptuik solution.)  The same critical solution is
approached in particular for initial data with $\Pi=0$ and hence no
angular momentum, and initial data where $\Pi=\bar\Phi$ and hence with
large angular momentum.  A scaling exponent of $\gamma\simeq 0.11$ is
observed in the Ricci scalar in subcritical evolutions. The critical
solution is purely real (up to an initial data-dependent constant
phase) and hence has no angular momentum. Only $m=1$ was investigated,
but it is plausible that a different critical solution exists for each
integer $m$.

Far from the black hole threshold, $J\sim M^2$ in the final black
hole, but nearer the black hole threshold, $J\sim M^6$, where $J$ and
$M$ are measured on the apparent horizon when it first
forms. $J/M^2\to 0$ is compatible with a non-rotating critical
solution.

In the absence of angular momentum, the wave equations for the real
and imaginary part of $\Phi$ decouple. Assuming that the background
critical solution is purely real with $\delta=0$ and $A\sim 1$, and
angular momentum is provided by a perturbation with $\delta\sim
e^{\lambda\tau}$, one would expect $\rho\sim|\nabla A|^2\sim
e^{2\tau}$ and $j\sim A|\nabla\delta| \sim
e^{(1+\lambda)\tau}$. Integrating over a region of size $e^{-3\tau}$
when the black hole forms, we find $M\sim e^{-\tau_*}$ and $J\sim
e^{(-2+\lambda)\tau_*}$. Then $J\sim M^6$ would imply $\lambda= -4$.

Lai \cite{Lai} has studied type I critical phenomena for boson
(massive complex scalar field) stars in axisymmetry, the first study
of type I in axisymmetry. He finds that the subcritical endstate is a
boson star with a large amplitude fundamental mode oscillation.


\subsection{Neutron star collision in axisymmetry}

A first investigation of type I critical collapse in an
astrophysically motivated scenario was carried out in
\cite{JinSuen}. The matter is a perfect fluid with ``Gamma law''
equation of state $P=(\gamma-1)\rho\epsilon$, where $\Gamma\simeq 2$
is a constant, $P$ is the pressure, $\rho$ the rest mass density and
$\epsilon$ the internal energy per rest mass (so that
$\rho(1+\epsilon)$ is the total energy density). The initial data are
constructed with $P=k\rho^\Gamma$ for a constant $k$, which
corresponds to the ``cold'' (constant entropy) limit of the Gamma law
equation of state. The initial data correspond to two identical stars
which have fallen from infinity. (The evolution starts at finite
distance, with an initial velocity calculated in the first
post-Newtonian approximation). The entire solution is axisymmetric
with an additional reflection symmetry that maps one star to the other. 

The parameter of the initial data that is varied is the mass of the
two stars. Supercritical data form a single black hole, while
subcritical data form a single star. The diagnostics given are plots
against time of the lapse and the Ricci scalar at the symmetry centre
of the spacetime, and of outgoing $l=2$ gravitational waves. Collapse
of the lapse and blowup of the Ricci scalar are taken as indications
of black hole formation. The limited numerical evidence is
compatible with type I critical phenomena, with the putative critical
solution showing oscillations with the same period in the lapse and
Ricci scalar. For the critical solution to be exactly time-periodic,
it would have to be spherical in order to not lose energy through
gravitational waves, and there is some evidence that indeed it does
not radiate gravitational waves.

Other one-parameter families of initial data were obtained by fixing
the mass of the stars in the first family very near its critical value
and then varying one of the following: the initial separation, the
initial speed, and $\Gamma$. For all approximately the same scaling
law for the survival time of the critical solution was found. However,
as all these initial data are very close together, this only confirms
the validity of the general perturbation theory explanation of
critical phenomena, but does not provide evidence of universality.

A key question that has not been answered is how shock heating affects
the standard critical collapse scenario. A priori the existence of a
universal spherical critical solution is unlikely in the presence of
shock heating as there is no dynamical mechanism to arrive at a
universal spherical temperature profile, in contrast to
non-dissipative matter models or vacuum gravity where all equations
are wave-like.


\subsection{Black hole collisions}

Interesting numerical evidence for critical phenomena in the glancing
collision of two black holes has been found by Pretorius
\cite{Pretorius07}. He evolved initial data for two equal mass
nonrotating black holes, with a reflection symmetry through
the orbital plane, parameterised by their relative boost at constant
impact parameter. The threshold in initial data space is between data
which merge immmediately and those which do not (although they will
merge later for initial data which are bound). The critical solution
is a circular orbit that loses $1-1.5\%$ of total energy per orbit
through radiation. On both sides of the threshold, the number of
orbits scales as
\begin{equation}
\label{Pretoriusn}
n\simeq -\gamma\ln|p-p_*|,
\end{equation}
for $\gamma\simeq 0.31-0.38$. The simulations are currently limited by
numerical accuracy to $n<5$ and $\Delta E/E<0.08$ but Pretorius conjectures
that the total energy loss and hence the number of orbits is limited only
by the irreducible mass of the initial data, a much larger number. In
particular, he speculates that for highly boosted initial data such
that the total energy of the initial data is dominated by kinetic
energy, almost all the energy can be converted into gravitational
radiation. 

The standard dynamical systems picture of critical collapse, with the
critical solution an attractor in the threshold hypersurface, appears
to be consistent with these observation. Pretorius compares his data
with the unstable circular geodesics in the spacetime of the
hypothetical rotating black hole that would result if merger occurred
promptly. These give orbital periods, and their linear perturbations
give a critical exponent, in rough agreement with the numerical values
for the full black hole collision. 

Pretorius does not comment on the nature of the critical solution, but
because of the mass loss through gravitational radiation it cannot be
strictly stationary. The mass loss would be compatible with
self-similarity, with a helical homothetic vector field, but exact
self-similarity would be compatible with the presence of black holes
only in the infinite boost limit. Therefore the critical solution is
likely to be more complicated, and can perhaps be written as an
expansion with an exactly stationary or homothetic spacetime as the
leading term.

Pretorius also
speculates that these phenomena generalise to generic initial data
with unequal masses and black hole spins which are not aligned. This
seems uncertain, given the claim by Levin \cite{Levin} that the
threshold of immediate merger is fractal if the spins are not aligned,
and that the system is therefore chaotic. However, Levin's analysis is
based on a 2nd order post-Newtonian approximation to general
relativity, in which there is no radiation reaction, while the rapid
energy loss observed here may suppress chaos. Nevertheless, the phase
space is much bigger when the orbit is not confined to an orbital
plane, and so the critical solution observed here may not be an attractor
in the full critical surface.


\subsection*{Thanks}

We would like to thank David Garfinkle for a critical reading of the
manuscript and Louisiana State University for hospitality while
this work was begun. JMM was supported by the I3P framework
of CSIC and the European Social Fund, and by the Spanish MEC Project
FIS2005-05736-C03-02.


\bibliography{critreview4.bbl}

\end{document}